\documentclass{article}

\usepackage{PRIMEarxiv}

\usepackage[utf8]{inputenc} 
\usepackage[T1]{fontenc}    
\usepackage{hyperref}       
\usepackage{url}            
\usepackage{booktabs}       
\usepackage{amsfonts}       
\usepackage{nicefrac}       
\usepackage{microtype}      
\usepackage{lipsum}
\usepackage{fancyhdr}       
\usepackage{graphicx}       
\usepackage{algorithm}
\usepackage{longtable}
\usepackage{pifont}
\usepackage{algpseudocode}
\graphicspath{{media/}}     
\usepackage{listings}

\usepackage{xcolor}
\definecolor{codegreen}{rgb}{0,0.6,0}
\definecolor{codegray}{rgb}{0.5,0.5,0.5}
\definecolor{codepurple}{rgb}{0.58,0,0.82}
\definecolor{backcolour}{rgb}{0.95,0.95,0.92}
\lstdefinestyle{mystyle}{
    backgroundcolor=\color{backcolour},
    commentstyle=\color{codegreen},
    keywordstyle=\color{blue},
    numberstyle=\tiny\color{codegray},
    stringstyle=\color{codepurple},
    basicstyle=\ttfamily\small,
    breakatwhitespace=false,
    breaklines=true,
    captionpos=b,
    keepspaces=true,
    numbers=left,
    numbersep=5pt,
    showspaces=false,
    showstringspaces=false,
    showtabs=false,
    tabsize=2
}
\title{Deploying Large Language Models with
Retrieval Augmented Generation
}

\author{
  Sonal Prabhune and Donald J. Berndt \\
  University of South Florida \\
  Tampa, Florida, USA\\
  \texttt{\{saprabhune@usf.edu, dberndt@usf.edu\}} \\
}

\begin{document}
\maketitle

\begin{abstract}
    Knowing that the generative capabilities of large language models (LLM) are sometimes hampered by tendencies to hallucinate or create non-factual responses, researchers have increasingly focused on methods to ground generated outputs in factual data. Retrieval-Augmented Generation (RAG) has emerged as a key approach for integrating knowledge from data sources outside of the LLM’s training set, including proprietary and up-to-date information. While many research papers explore various RAG strategies, their true efficacy is tested in real-world applications with actual data. The journey from conceiving an idea to actualizing it in the real world is a lengthy process. We present insights from the development and field-testing of a pilot project that integrates LLMs with RAG for information retrieval. Additionally, we examine the impacts on the information value chain, encompassing people, processes, and technology. Our aim is to identify the opportunities and challenges of implementing this emerging technology, particularly within the context of behavioral research in the information systems (IS) field. The contributions of this work include the development of best practices and recommendations for adopting this promising technology while ensuring compliance with industry regulations through a proposed AI governance model.
\end{abstract}


\section{Introduction}
     Recently, the landscape of artificial intelligence (AI) has changed drastically. With the emergence of large language models and their capabilities of generating textual responses based on various prompting techniques. A recent article by \textit{Bloomberg Intelligence} (BI) \cite{Bloomberg2023} on the explosive growth of generative AI in the next decade predicts, ``With the influx of consumer generative AI programs like Google’s \textsc{Bard} and OpenAI’s \textsc{ChatGPT}, the generative AI market is poised to explode, growing to \$1.3 trillion over the next 10 years from a market size of just \$40 billion in 2022.'' As we develop cutting-edge applications that use large language models, we must explore the various possibilities, understand the shortcomings, and develop ways to circumvent any obstacles to ensure a reliable human-artificial intelligence system. 
 
 There are different approaches, even within the narrow scope of implementing a solution, that use a large language model (LLM) to retrieve information from proprietary unstructured documents. Current approaches for retrieving information from unstructured text include fine-tuning an LLM, parameter efficient fine-tuning approaches such as \textsc{LoRA} \cite{hu2021lora}, prefix tuning \cite{liu2021p}, prompt tuning \cite{lester2021power}, p-tuning \cite{liu2023gpt}, various approaches in prompt engineering such as zero-shot \cite{kojima2022large}, few-shot \cite{parnami2022learning}, chain-of-thought \cite{wei2022chain} and \textsc{ReAct} \cite{yao2022react}. It is essential to understand the business case and apply appropriate approaches, often requiring experimentation such as prototyping and evaluating what works best. Various libraries and tools have emerged to support the fast development of LLM systems based on information retrieval from proprietary data. \cite{Chase_LangChain_2022} \cite{Liu_LlamaIndex_2022} \cite{Microsoft} 
 These tools enable the LLMs to behave as agents, \cite{wu2023autogen} function as reasoning engines, and invoke various tools supplied to the LLM to generate the required outcome. In addition, the LLM itself can be used to evaluate the efficacy of these generated outcomes. What remains for those implementing the Retrieval Augmented Generation (RAG) system is to make the right choices of these tools and to design a system that is most effective for their particular use cases. As simple as this might seem, the complexity arises from needing to implement such a system at scale while conforming to the industry's regulatory standards.
 
 As with any emerging technology, there is a lag between advancements in large language models (LLMs) with Retrieval-Augmented Generation (RAG), and their adoption and implementation in the industry.  We believe that valuable insights can be gained from the practical application of RAG with LLMs, which can help advance research and facilitate the next breakthroughs. This paper seeks to share these learnings and bridge the gap between the challenges encountered in the industry and the issues being tackled by researchers. We draw parallels between the challenges mentioned in a recent paper \cite{hevner2023research}, which discusses key aspects of human interactions with AI systems, and those we encountered in our field study. We suggest best practice recommendations for organizations venturing into the space of generative AI with RAG. We hope that this paper, detailing our challenges and insights, will not only serve as a valuable guide for organizations planning to implement RAG-based solutions using LLMs but also provide researchers with a deeper understanding of the real-world challenges faced in the industry when adopting new technology. 

The structure of this paper is as follows: We begin by exploring the research efforts aimed at enabling LLMs to adapt to domain-specific contexts, focusing on fine-tuning and prompt engineering. Following this, we delve into the work conducted with RAG to date. We then provide the background for our field study and detail the design of the pilot project, discussing the various challenges faced by the industry in developing such a system and the wide array of choices available for each component of the RAG system. Next, we outline the implementation of our pilot project, including the design choices and our approach to execution. We also address the importance and challenges of evaluating such a system, reviewing the different evaluation strategies, including those we employed. Our fieldwork insights, best practices, and the necessity of AI governance are discussed subsequently. Finally, we examine the implications of deploying these systems in production, with a particular focus on the challenges of human-artificial intelligence collaboration, to promote the development of robust, reliable, and compliant AI solutions across various industries. The paper concludes with a summary of our findings.

\section{Related Work}
\label{sec:headings}
    Prior to LLMs, the approaches to retrieval of information from unstructured text included statistical and machine learning-based approaches. Examples of statistical approaches include bag-of-words and term frequency-inverse document frequency (TF-IDF) \cite{ramos2003using}. For instance, the TF-IDF approach has been used for extracting ontological data from unstructured text \cite{berndt2010using}. On the other hand, machine learning approaches are based on generating word embeddings such as \textsc{Word2vec} \cite{mikolov2013efficient}, which captures the idea of similarity between words within documents.  

Recently, new technologies have emerged,
such as large language models that generate new text
and embeddings from existing documents.
Although very promising in their capabilities,
these models can suffer from hallucinations,
generating responses that are not grounded in facts.
LLMs come up with their responses mainly based on
next-word probability \cite{wolfram2023chatgpt}.
These hallucinatory tendencies impede the application of
LLMs to mainstream industry applications because of
obvious questions about the reliability and robustness of their responses
\cite{lewis2020retrieval} \cite{bang2023multitask}.
To address this, a lot of research has been published that suggests a variety of approaches which are summarized in table \ref{tab:long-lit}

\begin{figure}
\centering
\includegraphics[width=0.6\textwidth]{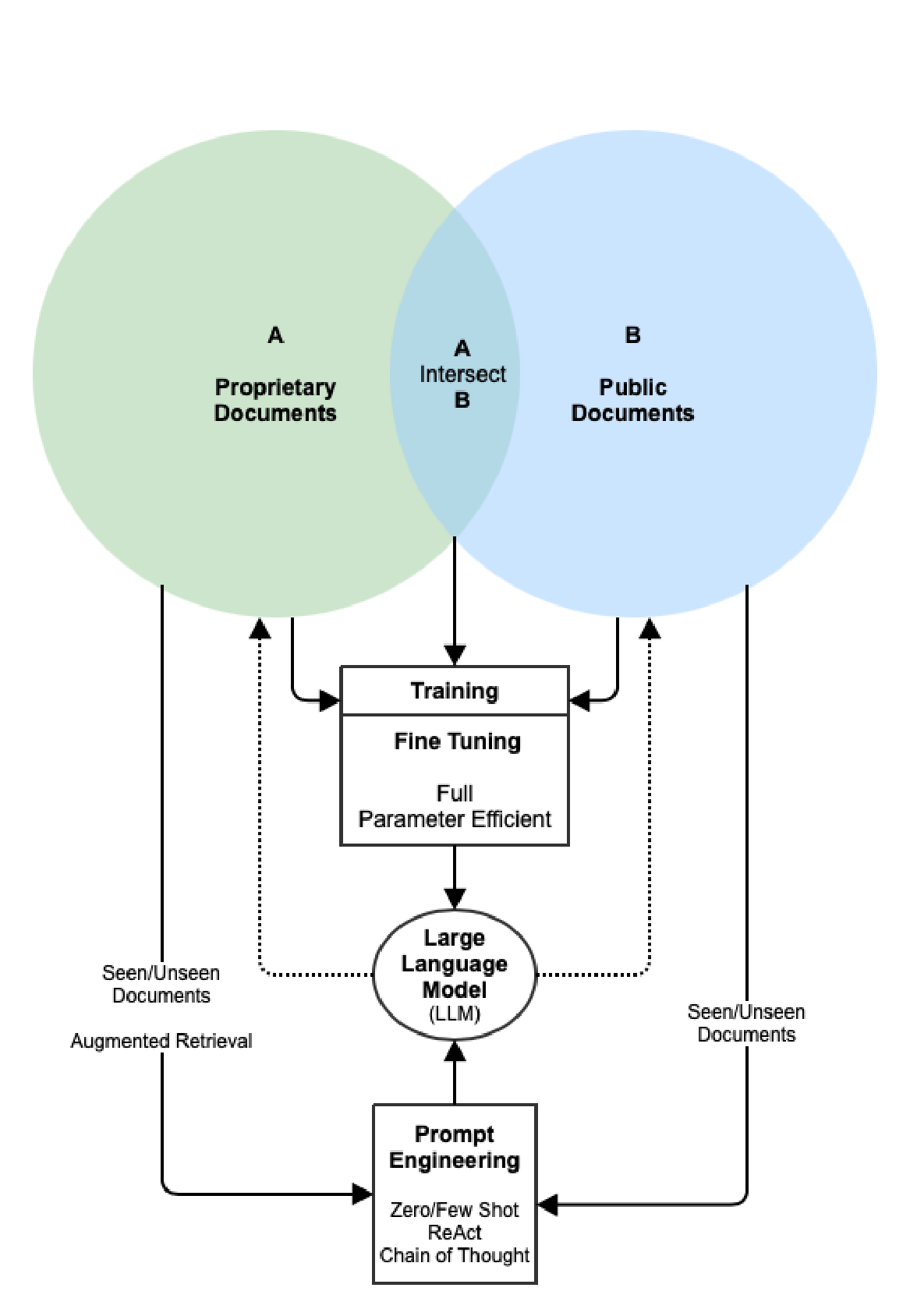}
\caption{Large Language Model Landscape}
\label{fig:llm}
\end{figure}

Figure \ref{fig:llm} provides an overview of the LLM landscape
highlighting the use of both proprietary ($A$) and public ($B$) documents.
An LLM is typically trained on vast amounts of public data
but can be customized by additional training on proprietary documents
(using full or parameter-efficient fine-tuning).
There could even be an overlap between proprietary and
public documents ($A \cap B$) if private documents are released
or public documents are retracted.
Of course, the documents generated by an LLM can be re-cycled
as training data as well.
Using an LLM involves prompt engineering with or without examples,
which can be drawn from seen or unseen documents.
Several possible approaches to
reduce hallucinations are outlined below.

\begin{itemize}
    \item \textbf{Full fine-tuning}:
    The pre-trained LLM understands the basic language construct,
    but it is not adapted for domain-specific context and vocabulary.
    This approach basically modifies the LLM's weights to
    adopt the context from the domain-specific dataset.
    It is one of the oldest approaches
    for domain-specific tuning of an LLM.

    \item \textbf{Parameter-efficient fine-tuning}:
    In this approach, the LLM is fine-tuned by modifying
    only a small number of parameters \cite{hu2021lora} \cite{lester2021power} rather than all of them,
    as in the case of full fine-tuning.
    This approach reduces the load with respect to
    computations, memory use and execution time.

    \item \textbf{Prompt engineering}:
    Prompt engineering is an approach where the LLM
    is given instructions to generate a certain response.
    Different approaches have been shown to enhance
    the quality of responses generated using prompt engineering.
    Some of them include zero-shot prompting \cite{kojima2022large}, 
    where the model is given only instructions but no examples,
    few-shot prompting where the model is given instructions and a few examples \cite{parnami2022learning}, 
    chain-of-thought prompting \cite{wei2022chain} 
    where the model is guided through steps for problem-solving,
    \textsc{ReAct} \cite{yao2022react} where the model is guided using reasoning
    and task-specific actions in an interleaved manner,
    or some combination of these different approaches.
\end{itemize}

\subsection{Retrieval Augmented Generation}
Retrieval-Augmented Generation (RAG) is an approach that enhances the capabilities of Large Language Models (LLMs) by augmenting their prompts with information retrieved from a knowledge base that the LLM has not seen during training. This method adds domain-specific context to the generated responses, leveraging the LLM's ability to generate text while incorporating embeddings from existing documents stored in vector databases. These databases are specifically designed to store, query, and retrieve sentence embeddings, making them integral to the RAG approach.

The field of RAG has rapidly evolved, as highlighted by various studies such as those surveying various RAG methodologies \cite{li2022survey}, \cite{zhao2024retrieval}, \cite{yu2024evaluation}, \cite{gao2023retrieval}, \cite{wu2024retrieval} and exploring advanced implementations like \textsc{RAG-end2end} \cite{siriwardhana2023improving}.

The table \ref{tab:long-lit} summarizes various research efforts in the field of Retrieval-Augmented Generation (RAG), each addressing different aspects and challenges of integrating retrieval mechanisms with large language models (LLMs) as well as the evaluation challenges of RAG systems.  Key studies include the foundational work \cite{lewis2020retrieval} that showed how incorporating retrieval into individual tasks with a single retrieval-based architecture is capable of achieving strong performance across several tasks. The survey \cite{li2022survey} highlights RAG's applications across multiple NLP tasks but notes areas needing improvement, such as optimization and multi-modalities. Other surveys \cite{zhao2024retrieval}, \cite{wu2024retrieval}, and \cite{gao2023retrieval} further categorize and review RAG techniques, discussing the limitations and future directions for RAG implementation in the industry, as well as discussing different RAG paradigms, including their application to various data stores and knowledge graphs. 
Practical evaluations and benchmarking are addressed in studies  \cite{salemi2024evaluating} and \cite{chen2024benchmarking}, which propose new methods for assessing RAG performance and robustness. Various studies propose performance improvement approaches, including \cite{siriwardhana2023improving}, which explores advanced implementations like \textsc{RAG-end2end}, \cite{wang2024searching}, which proposes a “Hybrid with HyDE” method for retrieval, and \cite{jiang2023active} proposes a Forward-Looking Active REtrieval augmented generation (FLARE). Some studies \cite{yu2022retrieval} provide solutions on RAG methods across heterogeneous knowledge. Others explore improved evaluation approaches such as \cite{yan2024corrective}, which proposes a Corrective Retrieval Augmented Generation (CRAG), a retrieval evaluator that can be coupled with various RAG approaches to improve the robustness of generation. These studies underscore the growing academic interest in RAG.

\begin{longtable}{ | p{0.18\linewidth} | p{0.2\linewidth} | p{0.22\linewidth} | p{0.15\linewidth} | p{0.15\linewidth} |}
\caption{Review of Researches Exploring Biases in Generative AI - Details by Research Paper} \label{tab:long-lit} \\

\hline \multicolumn{1}{|c|}{\textbf{Research Paper}} & \multicolumn{1}{c|}{\textbf{Summary}} & \multicolumn{1}{c|}{\textbf{Main Findings}} & \multicolumn{1}{c|}{\textbf{Models}} & \multicolumn{1}{c|}{\textbf{Retrieval}} \\ \hline \hline 
\endfirsthead

\hline \multicolumn{1}{|c|}{\textbf{Research Paper}} & \multicolumn{1}{c|}{\textbf{Summary}} & \multicolumn{1}{c|}{\textbf{Main Findings}} & \multicolumn{1}{c|}{\textbf{Models}} & \multicolumn{1}{c|}{\textbf{Retrieval}} \\ \hline \hline 
\endhead

\multicolumn{5}{r}{{Continued on next page}} \\ 
\endfoot

\hline 
\endlastfoot
\hline

Retrieval-Augmented Generation for Knowledge-Intensive NLP Tasks \cite{lewis2020retrieval} & This paper introduces general-purpose fine-tuning for RAG models by combining pre-trained parametric and non-parametric memory, specifically integrating a pre-trained seq2seq model with a neural retriever. & The study finds that RAG models produce more specific, diverse, and factual language than a parametric-only seq2seq baseline, reducing "hallucinations" by grounding generated text in factual data. & BART, BERT, RAG & DPR \\ 
\hline
A Survey on Retrieval-Augmented Text Generation \cite{li2022survey} & This survey summarizes key components of retrieval-augmented text generation, including retrieval metrics, sources, integration paradigms, and applications like dialogue response generation and machine translation. & RAG benefits dialogue systems, machine translation, language modeling, and more, but still needs improvement in retrieval sensitivity, optimization, and multi-modalities. & NA & NA \\ 
\hline
Retrieval-Augmented Generation for AI-Generated Content: A Survey \cite{zhao2024retrieval} & This survey categorizes RAG by augmentation methods for retrievers and generators, discusses RAG applications across various modalities and tasks, offering references for researchers and practitioners. & This survey categorizes RAG based on the different methodologies and applications providing diagrammatic representations. It discusses the limitations and provides future directions. & NA & NA \\ 
\hline
Evaluation of Retrieval-Augmented Generation: A Survey \cite{yu2024evaluation} & This survey addresses RAG evaluation challenges and proposes a Unified Evaluation Process focused on targets, datasets, and measures. & The analysis highlights the need for benchmarks that link retrieval accuracy with generative quality, considering real-world applications. & NA & NA \\ 
\hline
Retrieval-Augmented Generation for Large Language Models: A Survey \cite{gao2023retrieval} & This survey explores RAG paradigms (naive, advanced, and modular RAG), key technologies in retrieval, generation, and augmentation, and introduces evaluation frameworks and benchmarks.  & The survey discusses optimization strategies and retrieval types (iterative, recursive, and adaptive), and explores future prospects for scaling RAG for production-ready systems. & NA & NA \\ 
\hline
Retrieval-Augmented Generation for Natural Language Processing: A Survey \cite{wu2024retrieval} & This survey reviews different techniques of RAG, especially in the retriever and the retrieval fusions. & Provides algorithms for implementing and discusses the RAG training for the application of RAG in representative natural language processing tasks
and industrial scenarios & NA & NA \\ 
\hline
Improving the domain adaptation of retrieval augmented generation ({RAG}) models for open domain question answering \cite{siriwardhana2023improving} & 
This paper introduces RAG-end2end, which uses an auxiliary training signal to incorporate domain-specific knowledge and evaluates it across datasets from three domains. & The RAG-end2end model shows better performance than the original RAG and improves DPR performance more effectively than finetuning. & pre-trained BART & DPR (Pretrained dense retriever) uses 2 BERT models, FAISS indexing \\ 
\hline
Evaluating Retrieval Quality in Retrieval-Augmented Generation \cite{salemi2024evaluating} & This study presents eRAG, a method where each document in the retrieval list is individually used by the LLM, and the output is evaluated against ground truth labels for downstream tasks. & Experiments show that eRAG improves correlation with downstream RAG performance and Kendall’s tau, while also being up to 50 times more efficient in GPU memory usage compared to end-to-end evaluations. & T5-small with Fusion-in-Decoder (FiD) as LLM, Mistral for evaluation of retrieved documents & BM25, Faiss \\ 
\hline
Benchmarking Large Language Models in Retrieval-Augmented Generation \cite{chen2024benchmarking} & This study evaluates LLMs on noise robustness, negative rejection, information integration, and counterfactual robustness for RAG. & Findings indicate that LLMs exhibit a certain degree of noise robustness, struggle in terms of negative rejection, information integration, and dealing with false information. & ChatGPT-3.5-turbo, ChatGLM-6B, ChatGLM2-6B, Vicuna-7B-v1.3, Qwen-7B-Chat, BELLE-7B-2M & Google's API for web search and an open-source dense retriever (not specified) \\ 
\hline
Searching for Best Practices in Retrieval-Augmented Generation \cite{wang2024searching} & This study evaluates various solutions for RAG modules and recommends the most effective approach for each.  & he study finds that the best performance is achieved using the “Hybrid with HyDE” method for retrieval, monoT5 for reranking, Reverse for repacking, and Recomp for summarization.  & GPT-3.5-turbo-0125, GPT-3.5-turbo-instruct, Zephyr-7b-alpha, monoT5, monoBERT, TILDEv2,  RankLLaMA etc. & Approaches - HyDE, Hybrid Search, databases - Weaviate, Faiss, Chroma, Qdrant, Milvus, etc. \\ 
\hline
Active Retrieval Augmented Generation \cite{jiang2023active} & This paper introduces FLARE, which generates a temporary next sentence to retrieve relevant documents based on low-probability tokens. & Findings show that FLARE performs better than single-time retrieval on four benchmarked datasets & Any LLM but they used gpt-3.5-turbo, GPT-3.5 text-davinci-003 & FLARE \\ 
\hline
Retrieval-augmented Generation across Heterogeneous Knowledge \cite{yu2022retrieval} & This study provides solutions on RAG
methods across heterogeneous knowledge & Solutions like homogenizing different knowledge, multi-virtual hops retrieval, and reasoning over structured knowledge for RAG across heterogeneous data. & T5 & Dense retriever model DPR \\ 
\hline
Corrective Retrieval Augmented Generation \cite{yan2024corrective} & This study introduces Corrective Retrieval Augmented Generation (CRAG) as a retrieval evaluator to enhance the robustness of RAG approaches. & Experiments on four datasets show that CRAG significantly enhances RAG-based performance in both short- and long-form tasks. & s LLaMA2-chat13B, ChatGPT, CoVE65B & LLaMA2-7B,13B, Alpaca-7B,13B, SAIL, T5-based Retrieval Evaluator\\ 
\hline

\end{longtable}

Despite the extensive research, there is a noticeable lack of field studies and industry-shared experiences that could bridge the gap between theoretical research and practical application. For RAG systems to fully realize their potential, research must be conducted in tandem with industry implementations. This collaboration would help address real-world challenges, such as managing heterogeneous data sources, integrating separate knowledge bases, and optimizing system performance for production environments. Sharing field studies and insights from industry implementations can significantly enhance research, enabling more robust solutions that cater to the complex demands of real-world applications.

Given the complexity of real-world applications, particularly in industry settings that involve intricate socio-technical systems, more detailed research is needed to study the implementation and deployment of RAG systems in the field as pilot projects in the industry and their maturation into production. Such research should assess the maturity of this approach, identify and address process gaps, and generate ideas to enhance the reliability and adoption of RAG systems. Collaboration between industry and academia is, therefore, essential, as sharing the challenges faced in the field can significantly advance research on LLMs and RAG. Furthermore, it is crucial to establish best practices based on industry learning and to create a governance framework that ensures compliance with regulatory requirements when deploying RAG technologies.

\section{Design}
    The advent of large language models (LLMs) like \textsc{ChatGPT} has revolutionized the way information can be retrieved from proprietary documents. However, given the relative immaturity of these models, their integration into socio-technical systems must be meticulously designed to ensure that the final solution is both reliable and accurate. This design process involves extensive experimentation with prompt engineering, innovative strategies for data chunking, indexing, and retrieval, as well as iterative and thorough testing of the entire system.

In this section, we explore the challenges and opportunities of implementing RAG systems in production. We outline the key technological decisions required when designing such a system, focusing on the various components of RAG. We begin by discussing the need for and background of developing an RAG-based system for information retrieval in our field study, highlighting the main pain points in the existing system and the areas where RAG with LLMs was evaluated to add value. We then detail our design process, emphasizing our approach in light of user readiness for this human-AI system.

\subsection{Domain Background}

In our field study, focused on developing a pilot solution for an IT product company, the challenge of integrating LLMs like \textsc{ChatGPT} into socio-technical systems became particularly relevant. The variability of product versions across different customer environments in IT companies demands a trained support team capable of delivering prompt and accurate assistance. Addressing this need required a carefully designed system involving extensive experimentation with prompt engineering, innovative data chunking, and indexing methods, as well as iterative testing and refinement to ensure the reliability and accuracy of the LLM-based solution.

In IT organizations, support is often structured in multiple tiers, including internal and customer-facing support. Although the initial design of our system was not meant for direct customer access, it was essential to develop it in a way that allows customer support teams to efficiently retrieve the necessary information to address customer inquiries and escalations. As the system's reliability and accuracy are validated, there may be potential to explore its direct use by end customers.

\begin{figure}
\includegraphics[width=\textwidth]{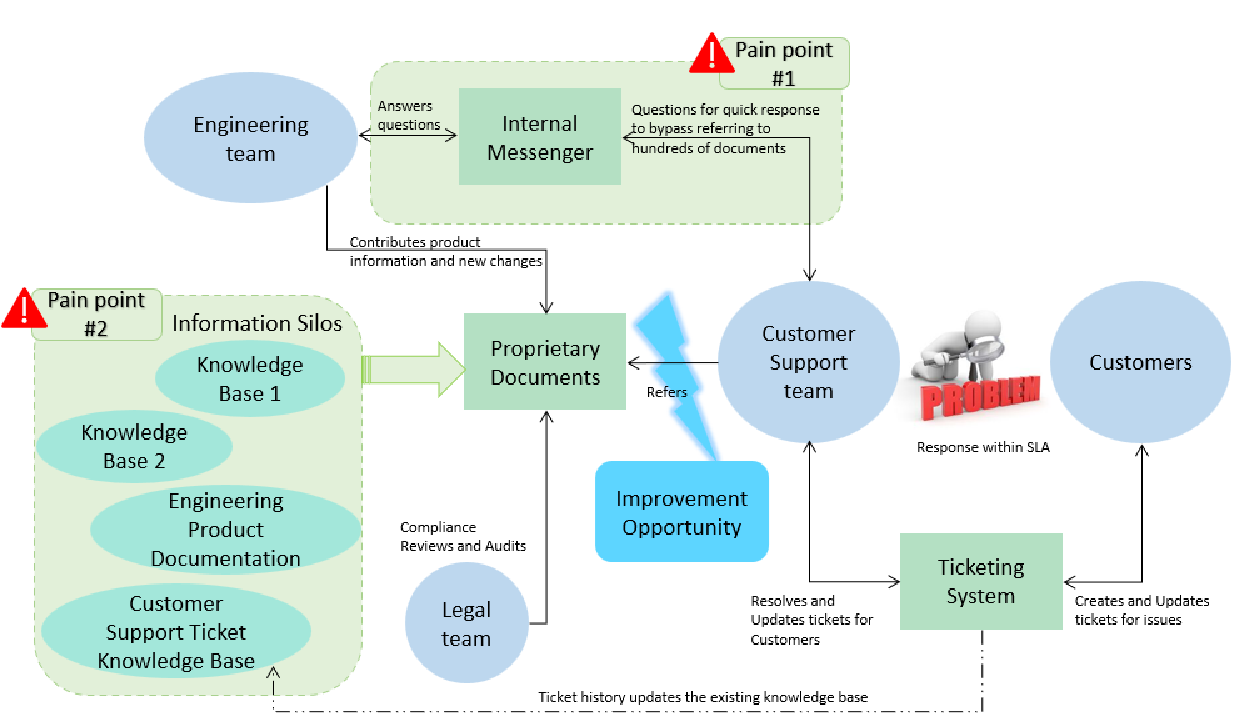}
\caption{Domain Background Analysis - Identification of Problem areas in the space} \label{fig:domain}
\end{figure}

In Figure \ref{fig:domain}, we identify key problem areas in the current state. 

The first pain point is the frequent back-and-forth communication between the customer support team and the engineering team, which is time-consuming and disruptive. This challenge stems from the need for customer support teams to meet Service Level Agreements (SLAs) by providing timely and accurate responses to customer tickets. Ensuring the accuracy of responses requires verification across various product documents, which differ by product and version. By improving the efficiency of information retrieval from these documents, these pain points can be mitigated, enabling support teams to meet SLAs more effectively.

The second pain point is the fragmented nature of information sources within most organizations. Information is often spread across multiple knowledge management systems, ticketing systems, databases, and other formats, each in its own silo. The integration of information across these disparate systems is complex and presents a significant challenge.

\subsection{Challenges and Opportunities}

In most organizations, retrieving information is not just about obtaining data; it’s about doing so quickly, accurately, and in a user-friendly manner. The ideal solution would allow for natural language queries without requiring users to have expertise in complex querying languages like SQL. Additionally, the information retrieved should be easy to understand or even summarized for the user’s convenience.

A significant challenge lies in the evolution of knowledge management systems alongside the organization itself. This evolution often results in inconsistencies in data formats, with information being both structured and unstructured. Organizations typically have a wide variety of document formats, such as .doc, .docx, .pdf, .csv, .xls, .txt, and HTML, among others. Furthermore, this information is often scattered across multiple systems, including various knowledge management frameworks, ticketing systems, and databases, each siloed with its own format. Integrating this diverse information into a cohesive system is inherently complex and requires a nuanced approach.

Beyond these technical challenges, several external factors influence the adoption and implementation of emergent technologies like Large Language Models (LLMs):
\begin{itemize}
    \item \textbf{Legal and governance}:
    Many organizations are cautious about exposing proprietary documents to external LLMs due to internal legal and compliance requirements. This includes concerns about data security and the risks associated with hosting sensitive information outside the organization’s infrastructure or sharing it with third-party vendors.

    \item \textbf{Financing}:
    Budgetary constraints often dictate whether an organization can implement a solution in-house or must opt for outsourcing. While outsourcing might provide a ready-to-use solution, it often comes with recurring subscription costs that can strain organizational budgets. On the other hand, in-house development requires a team with specialized skills for ongoing development, maintenance, and support.
    
    \item \textbf{Volatility}:
    The landscape of LLM technologies is rapidly evolving, with new libraries and frameworks emerging frequently. This volatility presents challenges not only for in-house development but also when negotiating contracts with third-party vendors. Organizations must carefully consider the total cost of ownership, including the long-term costs associated with keeping the system up-to-date.
\end{itemize}

When developing a Retrieval-Augmented Generation (RAG) system, particularly when using Large Language Models (LLMs) with proprietary documents, organizations must establish clear functional, non-functional, and technical requirements. (Please refer to Appendix \ref{sec:funcnonfunc}). In our fieldwork, we began this process by understanding the specific needs and expectations of stakeholders, including end-users, developers, and business leaders. The functional requirements focus on the system’s core capabilities, such as natural language understanding, contextual awareness, and integration with existing systems. The non-functional requirements, on the other hand, emphasize performance aspects like scalability, security, and reliability, ensuring the system can operate efficiently under varying conditions. Technical requirements detail the necessary infrastructure, tools, and frameworks that support the system’s development and operation. Stakeholder requirements and acceptance criteria further refine the project by aligning the system’s features with user needs and setting clear benchmarks for success. Throughout this process, involving key teams—such as legal, compliance, engineering, customer support, and quality assurance—ensured that the system not only met technical and functional standards but also adhered to organizational policies and regulations.

During the initial phase, we conducted feasibility checks to define the scope of a Minimum Viable Product (MVP) and to plan a roadmap for incremental development. This approach is crucial as it ensures that the system design remains flexible enough to evolve over time, allowing for adjustments as the technology and organizational needs change. Feasibility analysis and careful planning are critical to the overall success of the project, ensuring that the final product meets both the technical and business requirements of the organization.

To ensure the solution met user needs, we assembled a cross-functional team to collaboratively design the system. This approach enabled users to actively participate not only in defining requirements and conducting final acceptance testing but also in the iterative development and evaluation phases. By involving users throughout the process, the design was continuously refined, aligning more closely with their expectations and operational realities.

\subsection{User Readiness}
To assess user readiness and the acceptance of the new generative AI-based Information Retrieval System, we distributed a pre-survey questionnaire among the pilot group of users selcted randomly across various departments. This survey was a key step in our process, as it aimed to capture the users' familiarity with generative AI, their expectations, potential concerns, and overall sentiment towards the new technology. (Please refer to Table \ref{tab:presurvey} for the pre-survey analysis summary)

\begin{table}[ht]
\centering
\caption{Pre-Survey Analysis Summary}
\label{tab:presurvey}
\begin{tabular}{|p{0.25\linewidth}|p{0.5\linewidth}|p{0.15\linewidth}|}
\hline
\textbf{Survey Aspect} & \textbf{Summary} & \textbf{Percentage} \\ \hline
\textbf{Departments} & The pilot users came from various departments within the organization, including engineering, customer support, and legal. & N/A \\ \hline
\textbf{Experience with Generative AI} & Users with prior experience in generative AI at various levels of expertise. & 54\% \\ \hline
\textbf{Perceived Benefits} & Users anticipated that generative AI would improve efficiency by providing accurate and relevant information, aiding in troubleshooting, and offering fast access to product-specific data. & N/A \\ \hline
\textbf{Unsure About Benefits} & Users who were unsure if AI can make them or their teams more efficient. & 8\% \\ \hline
\textbf{Cross-Functional Communication} & Users expected the AI to streamline communication, reducing the need for meetings and increasing efficiency. & N/A \\ \hline
\textbf{Unsure but Optimistic} & Users who were uncertain but optimistic about AI's impact on cross-functional communication. & 8\% \\ \hline
\textbf{General Enthusiasm} & Users who were enthusiastic about the application, believing it could unify the organization and increase transparency. & N/A \\ \hline
\textbf{Unsure of Expectations} & Users who were unsure of what to expect from the application. & 20\% \\ \hline
\textbf{Concerns} & Concerns included phrasing questions correctly, AI accuracy, data privacy, over-reliance on AI, and potential job displacement. & N/A \\ \hline
\textbf{No Concerns} & Users who had no concerns about using the AI application. & 41\% \\ \hline
\textbf{Data Needs} & Users wanted the AI to retrieve data from various enterprise systems and document formats. & N/A \\ \hline
\textbf{Uncertain About Data Needs} & Users who were uncertain about their data expectations. & 8\% \\ \hline
\textbf{Excitement and Gratitude} & Users who expressed excitement and gratitude for being part of the pilot program. & 25\% \\ \hline
\end{tabular}
\end{table}

\begin{itemize}
    \item \textbf{Experience with Generative AI:} The survey revealed that 54\% of users had prior experience with generative AI, though the level of expertise varied. This indicated a moderate level of familiarity within the user base, suggesting that additional training might be necessary to bring all users to a similar level of comfort.

    \item \textbf{Perceived Benefits:} Users highlighted several potential benefits of the AI system, including improved accuracy and relevance of retrieved information, faster access to data, and enhanced troubleshooting capabilities. There was also a strong expectation that the system would streamline cross-functional communication and reduce the need for frequent meetings. About 8\% of users, however, were unsure whether AI would significantly impact their efficiency, indicating a need for clear demonstrations of the system’s value.

    \item \textbf{Expectations:} Enthusiasm for the application was high, with some users anticipating that it would foster inclusivity, transparency, and collaboration across the organization. However, 20\% of respondents were uncertain about what to expect, reflecting a need for better communication and expectation management.

    \item \textbf{Concerns:} While many users were excited, there were notable concerns, including issues related to question phrasing, the accuracy and reliability of AI responses, data privacy and security, and the potential for over-reliance on AI. Some users also feared that such a tool could eventually replace human jobs. Notably, 41\% of respondents did not express any concerns, suggesting varying levels of apprehension about the new technology.

    \item \textbf{Data Needs:} The majority of users expressed a need for the system to retrieve data across various enterprise systems, in addition to documents in multiple formats. About 8\% of users were uncertain about what data the AI should provide, indicating a need for clearer guidelines on the system's capabilities.

    \item \textbf{General Sentiment:} Overall, the sentiment was positive, with 25\% of users expressing excitement and gratitude for being part of the pilot program. This enthusiasm will likely aid in the system’s adoption, provided that their concerns are adequately addressed through training and iterative improvements.
\end{itemize}

This pre-survey analysis underscores the importance of addressing user concerns and managing expectations while leveraging the enthusiasm and existing familiarity with generative AI to facilitate a smooth rollout of the new system. (Please refer to Appendix \ref{sec:presurvey} for details on our pre-survey questions and its analysis).

In the development of Retrieval-Augmented Generation (RAG) systems, selecting the right model, vector database, and cloud provider is critical to ensuring optimal performance and alignment with business objectives. The landscape of available technologies is vast, with each offering unique features, scalability options, and integration capabilities. For instance, different LLMs vary in their context windows, multimodal capabilities, and accessibility, influencing their suitability for specific use cases. Similarly, the choice of vector databases, directly impacts the system's ability to efficiently manage and search through large datasets. Furthermore, the cloud provider must be carefully chosen to support the desired scalability, security, and compliance needs. Each of these decisions must be tailored to the specific requirements of the business case, as the right combination can significantly enhance the effectiveness of the RAG system while ensuring it meets the operational demands and strategic goals of the organization. In the following subsections, we explore the available options for each of these key components of the RAG system.

\subsection{Model Selection Criteria}

\begin{table}[ht]
\centering
\caption{Comparison of Large Language Models (LLMs)}
\label{tab:LLM}
\begin{tabular}{|p{0.1\linewidth}|p{0.12\linewidth}|p{0.12\linewidth}|p{0.1\linewidth}|p{0.1\linewidth}|p{0.1\linewidth}|p{0.1\linewidth}|}
\hline
\textbf{LLM} & \textbf{Organization} & \textbf{Multimodal? } & \textbf{Access} & \textbf{Parameters} & \textbf{Token context window} & \textbf{Open Source} \\ \hline
GPT & OpenAI & \checkmark & Chatbot and API & 175B+ for (GPT3). Not Published for GPT4 & GPT-4 (32K version): Has a context length of 32,768 tokens & \ding{55} \\ \hline
Gemini & Google & \checkmark & Chatbot and API & Upto 27B Parameters & Upto 2M & \ding{55} \\ \hline
Llama3 & Meta & \ding{55} & Chatbot and open & Upto 405B & Customizable model to extend token window from 8k to > 1040K & \checkmark \\ \hline
Claude & Anthropic & \checkmark & Chatbot and API & Not Published & Upto 200,000 tokens & \ding{55} \\ \hline
\end{tabular}
\end{table}

The table \ref{tab:LLM} compares several prominent LLMs, including GPT from OpenAI, Gemini from Google, Llama from Meta, and Claude from Anthropic. It details key attributes such as whether each model supports multimodal inputs, its access methods, parameter sizes, token context windows, and open-source availability. For instance, GPT is not open-source. In contrast, Llama, developed by Meta, is open-source and supports up to 405 billion parameters but does not handle multimodal inputs. Gemini and Claude, from Google and Anthropic, respectively, support multimodal inputs but have varying access methods and token context capabilities. This comparison helps in selecting the most suitable LLM based on specific requirements such as model size, context window, and open-source needs.

When selecting a large language model (LLM) for deployment, it is essential to begin with a thorough evaluation of models that have been benchmarked on recognized datasets. Resources such as the Open LLM Leaderboard \cite{open-llm-leaderboard-v2} and survey research papers \cite{chang2024survey} provide valuable insights into the performance and suitability of different LLMs for specific use cases.

However, the selection process must also consider the organization's resource constraints and policies. Critical decisions include whether to use locally hosted models like LLaMA or public API models like ChatGPT, as well as whether to opt for open-source models or proprietary solutions. The size of the model and the choice between versions should be guided by the available infrastructure for hosting, the cost associated with token usage, and the overall billing rates.

While larger models generally excel at complex reasoning tasks, smaller models that have been fine-tuned can often deliver superior performance for specific tasks \cite{hoffmann2022training}. This makes model selection not just a technical decision but also one that involves strategic consideration of the trade-offs between performance, cost, and alignment with organizational goals.

\subsection{Vector Database Selection}

\begin{table}[ht]
\centering
\caption{Comparison of various Vector Stores}
\label{tab:vector}
\begin{tabular}
{|p{0.1\linewidth}|p{0.18\linewidth}|p{0.13\linewidth}|p{0.15\linewidth}|p{0.16\linewidth}|p{0.12\linewidth}|}
\hline
\textbf{Vector Store} & \textbf{Key Features} & \textbf{Use Cases} & \textbf{Scalability} & \textbf{Integration} & \textbf{Indexing} \\
\hline
Weaviate & Hybrid search (vector + keyword), Scalable, Flexible schema, Built-in ML model support, RESTful API / GraphQL & Semantic search, Knowledge graphs, Recommendation systems & High, supports distributed deployment & ML frameworks, REST API, GraphQL & Hybrid search capabilities \\
\hline
Faiss & High performance, Various indexing strategies, Customizable, Integrates with ML libraries & Similarity search, Nearest neighbor search, Clustering & High, optimized for large datasets & ML libraries, low-level control & Flat, IVFPQ, HNSW \\
\hline
Chroma & Real-time search, Scalable, Simple API, ML model integration & Real-time vector search, Recommendation engines, Personalization & High, designed for horizontal scaling & ML models, simple API & Optimized for real-time search \\
\hline
Qdrant & Distributed deployment, Advanced indexing, REST / gRPC APIs, Flexible data support & Similarity search, Recommendation systems, Personalization & High, supports distributed deployment & REST \& gRPC APIs, flexible integration & Advanced indexing techniques \\
\hline
Milvus & High performance, Distributed \& scalable, Multiple indexing options, Integration with data processing frameworks & Large-scale vector search, Machine learning, Data analysis & High, supports distributed deployment & Various data processing frameworks, ML libraries & IVF, HNSW, ANNOY \\
\hline
Pinecone & 	Managed serverless service
Automatic scaling
Compatible with embeddings from any AI model or LLM & Hybrid Search - vector search with keyword boosting
Recommendation engines
Large-scale vector search & High, managed serverless service with auto-scaling & Easy integration with popular ML frameworks, data sources, and models & Live index updates, Multiple indexing strategies tailored for performance \\
\hline
Elastic Relevance Search Engine &  Search Engine)	Full-text search
Relevance ranking
Hybrid search (vector + keyword)
Extensive ecosystem of plugins & Full-text search
E-commerce search
Hybrid search & High, supports distributed deployment and scaling & Extensive APIs
Supports a wide range of data types and ML integrations
Extensive ecosystem of plugins & Inverted index
k-NN search
Dense vector search through plugins \\
\hline
\hline
\end{tabular}
\label{tab:comparison}
\end{table}

The table \ref{tab:vector} provides a comparative overview of several vector stores—Weaviate, Faiss, Chroma, Qdrant, and Milvus—highlighting their key features, use cases, scalability, and integration capabilities. Each vector store offers distinct advantages tailored to specific needs in Retrieval-Augmented Generation (RAG) implementations. For instance, Weaviate excels in hybrid search with flexible schema support and built-in ML model integration, making it suitable for semantic search and knowledge graphs. Faiss, known for its high performance and various indexing strategies, is ideal for large datasets and similarity searches. Chroma stands out with real-time search capabilities and a simple API, which is advantageous for applications requiring real-time vector search and personalization. Qdrant offers advanced indexing and distributed deployment, supporting similarity search and recommendation systems. Milvus, with its high performance and multiple indexing options, is well-suited for large-scale vector searches and data analysis.

Selecting the right vector store is crucial for effectively implementing RAG systems in industry settings. The choice depends on factors such as the required scalability, the nature of the data, and integration needs with existing systems. For successful RAG deployment, organizations must match the vector store’s features with their specific use cases and operational requirements. For example, businesses needing robust distributed deployment and flexibility might opt for Weaviate or Qdrant, while those focusing on real-time search and personalization could benefit from Chroma. Understanding these differences ensures that the selected vector store aligns with the project’s goals, optimizing performance and efficiency in handling diverse data sources and retrieval tasks.

\subsection{Cloud Vendor Selection and AI Ops}
\begin{table}[ht]
\centering
\caption{Comparison of AI Cloud Platforms}
\label{tab:Cloud}
\begin{tabular}{|p{0.1\linewidth}|p{0.15\linewidth}|p{0.15\linewidth}|p{0.15\linewidth}|p{0.15\linewidth}|p{0.15\linewidth}|}
\hline
 & \textbf{AWS} & \textbf{Microsoft Azure} & \textbf{Google Cloud Platform (GCP)} & \textbf{IBM Cloud} & \textbf{Oracle Cloud} \\ 
\hline
\textbf{Service Breadth and Depth}     & Extensive AI/ML offerings for various use cases. & Extensive AI/ML offerings for various use cases. & Good in AI services like Vertex AI, Vision AI. & Watson offers strong AI services, especially NLP and ML.  & Growing AI offerings but less comprehensive. \\ \hline
\textbf{Ease of Use and Integration}   & Highly intuitive, strong integration with AWS services and other systems. & Highly intuitive, strong integration with Azure services and other systems. & Seamless with Google services; challenging with third-party. & Developing, good with IBM ecosystem. & Decent, but less user-friendly.   \\ \hline
\textbf{Performance and Scalability}   & Highly scalable AI-optimized infrastructure. & Strong scalability with specialized VMs. & Excellent performance in data-heavy tasks. & Limited scalability, expensive. & Decent, but less scalable. \\ \hline
\textbf{Cost Efficiency}               & Competitive pricing, expensive for large-scale AI. & Flexible pricing, competitive. & Cost-effective, especially with TPUs. & Cost-effective Watson services, but requires higher investment. & Competitive, but Oracle-centric. \\ \hline
\textbf{AI-Optimized Hardware}         & Advanced AI-platforms like Inferentia, Trainium, NVIDIA GPUs. & Wide range of AI-optimized VMs with NVIDIA GPUs. & Leading-edge TPUs for AI/ML workloads. & PowerAI with NVIDIA GPUs, evolving. & Solid hardware options, less innovation. \\ \hline
\textbf{Generative AI capability}      & Comprehensive AI services, including Amazon Bedrock, Amazon Q, Amazon Lex & Comprehensive AI services, access to OpenAI models (GT-4, Dall-E, Codex). & Comprehensive AI services, including image and music generation. & Limited models, focus on textual services. & Limited models, focus on chatbots, summarization. \\ \hline
\textbf{Support and Community}         & Large developer community, strong support. & Large developer community, strong support. & Large developer community, backed by Google AI research. & Strong support, smaller community. & Good support, smaller community. \\ \hline
\end{tabular}
\end{table}

The success of any AI application, especially those leveraging large language models, is closely tied to the infrastructure and services provided by major cloud providers. Studies \cite{van2024big} show that cloud giants like Amazon, Microsoft, and Google offer a comprehensive suite of AI and infrastructure services, including compute power, application development, security and compliance, and industry-specific solutions, underscoring the integral role of these cloud platforms in supporting AI development. These services are integral to the development, deployment, and scaling of AI applications.

Choosing the right cloud vendor is a critical decision that must be made early in the project. This decision impacts not only the development and deployment phases but also the long-term sustainability and scalability of the AI application. Enterprises must plan for the ongoing management of AI operations (AI Ops) within their broader cloud strategy, ensuring that AI applications can evolve and adapt in line with both technological advancements and business needs.

Table \ref{tab:Cloud} provides a comparative analysis of major AI cloud platforms—AWS, Microsoft Azure, Google Cloud Platform (GCP), IBM Cloud, and Oracle Cloud—across several key performance indicators (KPIs) relevant to AI operations (AI Ops). The table examines the service breadth and depth, ease of use and integration, performance and scalability, cost efficiency, AI-optimized hardware, generative AI capability, and the support and community offered by each platform. This comparison aims to highlight the strengths and limitations of each provider, helping users select the most appropriate cloud platform for their specific AI needs and operational contexts.

It is essential to ensure that the design process lays a robust foundation for developing a reliable and efficient system. By carefully selecting appropriate models, cloud infrastructure, and AI operations strategies, we align the system's architecture with the organization's technical requirements and user expectations. With the design principles firmly established, the next phase, the implementation phase, can focus on translating these design elements into a functional software prototype. The forthcoming Implementation section will detail the construction of the system, including workflow diagrams, prompt engineering, and semantic retrieval to enable the transformation of the design into a tangible, operational system.

\section{Implementation}
\begin{figure}
\includegraphics[width=\textwidth]{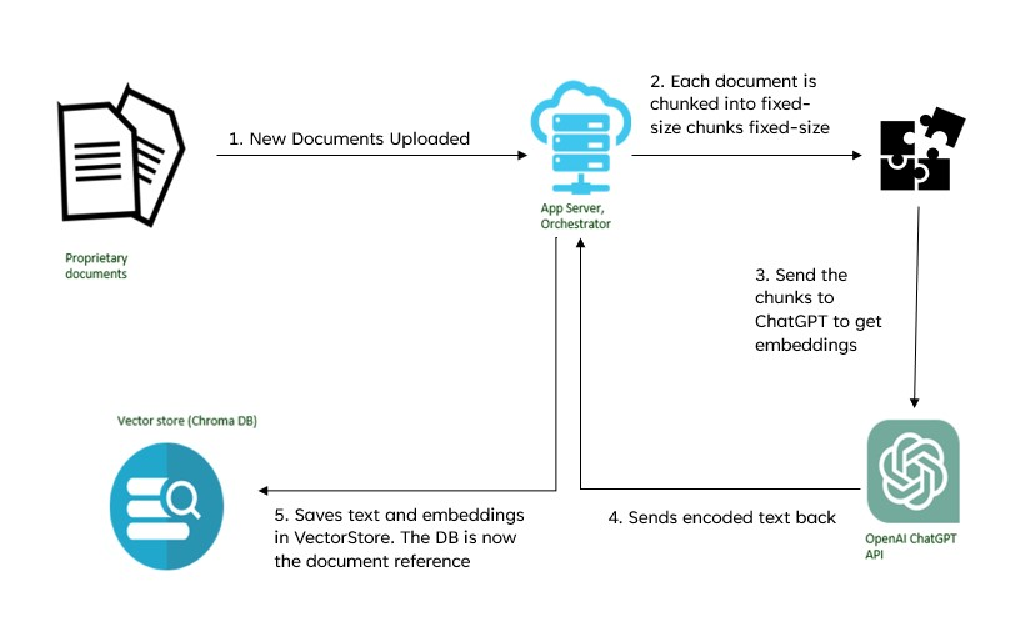}
\caption{Workflow for Document Pre-processing} \label{fig:doc}
\end{figure}

\begin{figure}
\includegraphics[width=\textwidth]{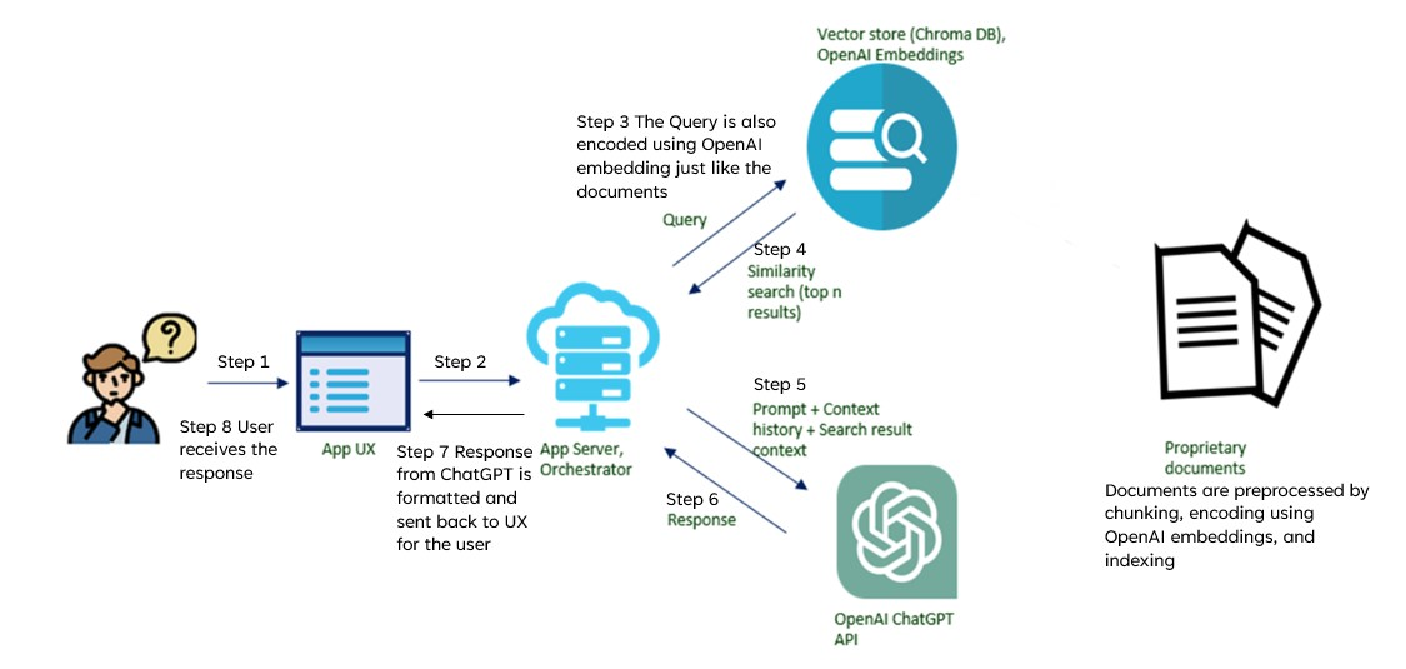}
\caption{System Operation Workflow} \label{fig:sys}
\end{figure}

    In the implementation phase, we transitioned from design to development, bringing the planned system architecture to life through a functional software prototype. This phase focused on translating the theoretical framework and design principles into a tangible system, ensuring that each component aligned with the specified requirements. We meticulously defined the system architecture, emphasizing the integration of various modules and the flow of information between them. The workflow, as depicted in Figures \ref{fig:doc} and \ref{fig:sys}, showcases the algorithm's overall structure, guiding the sequence of operations within the system. The subsequent sections provide a detailed breakdown of the implementation steps, highlighting key decisions, challenges encountered, and the strategies employed to achieve the desired functionality. 

\subsection{Workflow}

After outlining the system architecture and detailing the design principles, we focus on the operational workflow of the prototype. This workflow is central to the system's functionality, ensuring that each component interacts seamlessly with the others. The steps outlined below provide a clear path from data ingestion to the final user interaction, illustrating how the system processes, indexes, and retrieves information to meet user queries efficiently.

In the following workflow subsection, we break down the system's operations step by step, starting from how documents are read and processed to how user queries are handled and responses generated. This structured approach ensures that each stage of the process is clearly understood, enabling a smooth transition from design to fully functional implementation.

\begin{algorithm}
\caption{Algorithm for Retrieval Augmented Generation using LLM}\label{alg:rag-llm}
\begin{algorithmic}[1]

\State Create a prompt template with basic instructions to the LLM and include placeholders for retrieved document, user query and chat history
\State Instantiate splitter to chunk documents
\State docs = Split documents using splitter 

\State Instantiate embeddings = OpenAIEmbeddings()
\State Instantiate Vector Database Chroma to store the split documents "docs" and their embeddings 
\State Specify the vector database search type and number of top matching results for retrieval 

\State Instantiate retriever 

\State Save vector database in a persistent directory
   
\While{Chat session not exited}
    \State Get the query from the application's UI
    \State Generate Embeddings for the query
    \State Search top 3 results from the vector database that have embeddings "similar" to the query embeddings
    \State Update the prompt template with the query and retrieved document and the chat history
    \State Send the prompt to the LLM ChatGPT API
    \State Receive the response from the ChatGPT API 
    \State Save information on query, retrieved document titles and response in a log file
    \State Reformat the response for presentation to the user via UI
\EndWhile
\end{algorithmic}
\end{algorithm}

\begin{enumerate}
\item The PDF documents are read from a specific directory location.

\item They are split into chunks, with each chunk containing 1000 characters and overlap of 50 between chunks.
\item Next, these chunks are indexed. Below is the sample code, which shows one approach to implementing this using \textsc{LangChain} libraries. \cite{Chase_LangChain_2022}
\begin{lstlisting}[language=Python,
caption=Initializing splitter for chunking documents and also initializing embeddings model database and retriever.,
label=py:index]
from langchain.indexes import VectorstoreIndexCreator
from langchain.embeddings.openai import OpenAIEmbeddings
from langchain.text_splitter import CharacterTextSplitter
from langchain.vectorstores import Chroma

text_splitter = CharacterTextSplitter(chunk_size=1000,
                                      chunk_overlap=50)
docs = text_splitter.split_documents(documents)
embeddings = OpenAIEmbeddings()
db = Chroma.from_documents(documents=docs, embedding=embeddings)
ret = db.as_retriever(search_type="similarity", search_kwargs={"k":3})
\end{lstlisting}

\item After that, we create the embeddings. Since we are using \textsc{ChatGPT} as our large language model, we must use OpenAI’s embeddings. Below is the sample code, which shows one approach of implementing this.
\item The embeddings are stored in an instance of
the \textsc{Chroma} vector database.
\begin{lstlisting}[language=Python,
caption=Chunking Indexing and Storing the embeddings in the Chroma vector database and Initializing the retriever.,
label=py:chroma]
from langchain.vectorstores import Chroma
from langchain.indexes import VectorstoreIndexCreator
from langchain.embeddings.openai import OpenAIEmbeddings

embeddings = OpenAIEmbeddings()
index = VectorstoreIndexCreator(
    text_splitter=text_splitter,
    vectorstore_cls=Chroma
).from_loaders([loader])
persist_directory = '../../db' 
if (os.path.isdir(persist_directory)):
    vectordb = Chroma(
    persist_directory=persist_directory, 
    embedding_function=embeddings)
else:
    vectordb = createDB(
                docs=docs, 
                embeddings=embeddings, 
                persist_directory=persist_directory)
retriever = db.as_retriever(
    search_type="similarity", 
    search_kwargs={"k":3})
r = retriever.get_relevant_documents(q)[:top]
\end{lstlisting}

\item Steps 1 to 5 happen once in a separate application workflow. When a user uploads new documents, we repeat the process for those new documents.

\item The application has a React.js frontend and a Python Flask API backend. When a user asks the chatbot, for example, “What features are in the latest release of a particular product?” the question gets sent to the backend Python Flask API in a serialized JSON object along with the context history. Note that here, \textsc{ChatGPT} is expected to understand the meaning of the “latest” release. It would have to scan through all the documents to identify which release is the latest in this case.

\item When the question arrives at the backend, we first encode it using the same OpenAI embeddings used for encoding the document. This is necessary for retrieving similar documents from the vector store.

\item We run a semantic search on the vector store, \textsc{Chroma}, and return the top 3 matching documents. Below is the sample code, which shows one approach of implementing this.
\begin{lstlisting}[language=Python,
caption=Query the Chroma vector database for top 3 documents based on the user question.,
label=py:top-3]
query = "What is in the March release?"

retriever = vectordb.as_retriever(search_type="similarity",
                                  search_kwargs={"k":3})
r = retriever.get_relevant_documents(query)[:3]
\end{lstlisting}
\item We then embed these documents as ``sources'' in the prompt along with the question asked by the user and the context history and send it to the \textsc{ChatGPT} API. We are using the GPT-4-Turbo model \cite{ChatGPT4Turbo2024}.

\item When we get a response from \textsc{ChatGPT}, we send it back to the user interface (UI) for formatting and presentation. Below is a sample of the response with proprietary information redacted.
\begin{verbatim}
[Document
    (page_content='Summer Release 2022 Release Notes \n
    March 30, 2022 Release (Summer Release)\n
    The March 30, 2022 Release (Summer Release) 
    contains the following information.\nNew Features\n
    Inventory Management.\n
    New User Interface.\n
    User Management.\n
    Additional Opportunities.', 
    metadata={
        'source': '/home/MyApp/data/Mar_2022_Release_Notes.pdf', 'page': 10}), 
Document
    (page_content='Summer Release 2023 Release Notes\n 
    February 28, 2023 Release (Summer Release)\n
    The February 28, 2023 release 
    (Summer Release) contains the 
    following information.\n
    Enhancements\nSearch Enhancements.', 
    metadata={
        'source':'/home/MyApp/data/Feb_2023_Release_Notes.pdf', 
        'page': 7}), 
Document
    (page_content='Summer Release April 2022 Release Notes\n
    April 30, 2022 Release (Summer Release)\n
    The April 30, 2022 Release (Summer Release) 
    contains the following information.\nNew Features\n 
    Dashboard Updates.\n Enterprise Master.', 
    metadata={
        'source': '/home/MyApp/data/April_2022_Release_Notes.pdf', 'page': 9})]
\end{verbatim}

\end{enumerate}

\subsection{Prompt Engineering: Developing the Prompt}

Prompt engineering is how LLMs are programmed via prompts \cite{white2023prompt}. Prompts are natural language text written to instruct an LLM. There are various approaches to prompting, such as zero-shot, few-shot, chain-of-thought, and \textsc{ReAct} prompting \cite{kojima2022large} \cite{yao2022react} \cite{parnami2022learning}.

 Many of the libraries now provide prompt templates that could be used for generating standardized prompts. This helps in structuring and organizing your prompts as well as standardizing them for maintenance. Using certain instructions has been shown to reduce hallucinations \cite{tonmoy2024comprehensive} (for example, \textit{Answer ONLY with the facts listed in the sources below. If there isn't enough information below, say you don't know. DO NOT generate answers that don't use the sources below. If asking a clarifying question to the user would help, ask the question}). 
 To ensure security, we ensured that \textsc{ChatGPT} did not generate any code or SQL queries in their response (for example, \textit{Do NOT generate any code or SQL queries even when the user asks.}). Finally, to ensure multi-lingual questioning (\textit{If the question is not in English, translate the question to English before generating the search query.}). The chain-of-thought approach can be helpful in instructing the LLM in specific cases, such as searching for documents that contain release notes for products where the document name and structure are very similar between the different releases.
The approach is outlined below.

\begin{verbatim}
    prompt_prefix = """<|im_start|>system
    Assistant helps the company employees 
    with their product questions, and 
    questions about product releases. 
    Be brief in your answers.
    If asking a clarifying question to 
    the user would help, ask the question.
    Answer ONLY with the facts listed in 
    the list of sources below.
    Look at into all the sources. 
    If there isn't enough information 
    below, say you don't know. 
    Do not generate answers that don't 
    use the sources below. 
    For tabular information return it 
    as an HTML table. Do not return markdown format.
    Each source has a name followed by 
    colon and the actual information.
    Do not generate any code or 
    SQL statements in any format. 
    If prompted to generate code or 
    SQL queries say I am not allowed 
    to generate code or SQL queries.
    For questions about releases and 
    new features look at all the sources.
    {follow_up_questions_prompt}
    {injected_prompt}
    Sources:
    {sources}
    <|im_end|>
    {chat_history}
    """

    follow_up_questions_prompt_content = """Generate three 
    very brief follow-up questions that 
    the user would likely ask next about their products. 
    Use double angle brackets to reference 
    the questions, e.g. <<Could you please 
    clarify what exactly are you looking for?>>.
    Try not to repeat questions that have 
    already been asked.
    Only generate questions and do not 
    generate any text before or after the 
    questions, such as 'Next Questions'"""

    query_prompt_template = """Below is a 
    history of the conversation so far, 
    and a new question asked by the user 
    that needs to be answered by searching 
    in a knowledge base about products and releases.
    Generate a search query based on the 
    conversation and the new question. 
    Do not include cited source filenames 
    and document names e.g info.txt or 
    doc.pdf in the search query terms.
    Do not include any text inside [] or <<>> 
    in the search query terms.
    If the question is not in English, 
    translate the question to English 
    before generating the search query.

    Chat History:
    {chat_history}

    Question:
    {question}

    Search query:
    """
\end{verbatim}

\begin{enumerate}
\item We created a basic structure with all the system instructions, calling it the prompt template.
\item This template contained placeholders for “follow-up questions,” where the follow-up question would later be embedded. 
\item It also contained placeholders for “sources” where the retrieved documents would be embedded and placeholders for the chat history and user questions.
\item For contextual history, we had to introduce specific begin and end tags for the instructions in the prompt. We found that it was helpful to demarcate the prompt and the history. For example, consider these tags: \textit{$<|im\_start|> <|im\_end|>$}.
\item We also provided examples for follow-up questions, such as the following. \textit{What are the features of the latest release of this product?}
\end{enumerate}

\subsection{Logging}

Incorporating robust logging mechanisms is essential for the effective operation and optimization of LLM-based systems. Logging every input and output interaction with the LLM is not just a best practice for troubleshooting; it is also vital for fine-tuning the model over time. Tracking the number of tokens in each input, context, and response—whether the LLM is hosted locally or accessed via an API—provides valuable insights. This data is crucial for planning infrastructure requirements and ensuring the scalability of the system, particularly as traffic increases. Logging also aids in understanding the model's behavior, identifying patterns of errors or inefficiencies, and making informed decisions about future adjustments to the system architecture.

\subsection{Semantic Retrieval}

For the semantic search and retrieval component of our system, we conducted an extensive evaluation of various vector stores, both open-source and commercial, including \textsc{Chroma}, \textsc{Pinecone}, and \textsc{ElasticSearch}. A key finding from our evaluation was that misalignments between the generated responses and the user's queries often stemmed from issues in the retrieval process rather than the LLM itself. To address this, we implemented functionality to display links to the documents retrieved during the search. This approach not only provided users with citations for fact-checking but also allowed us to diagnose and refine our retrieval strategies when incorrect documents were retrieved. We also implement relevance-checking approaches, which we will discuss in the evaluation section.

Initially, we chunked documents without overlap, but further experimentation revealed that overlapping chunks improved retrieval accuracy. We also varied the number of retrieved results and tested different retrieval strategies, including similarity search, maximum marginal relevance, and hybrid approaches that combine keyword search with semantic retrieval. This iterative process helped us identify the optimal configuration for our specific documents and use case.

Recent research underscores the effectiveness of Retrieval-Augmented Generation (RAG) approaches, particularly when combining semantic retrieval from vector databases with other techniques like knowledge graphs or fine-tuning strategies. Knowledge graphs offer a complementary retrieval method, storing and querying information in the form of entity-attribute-attribute value triples. However, generating knowledge graphs from proprietary documents poses significant challenges, especially when dealing with inconsistently structured documents that have evolved over time within various departments of an organization.

Fine-tuning, including Parameter Efficient Fine-Tuning (PEFT), also presents challenges. While PEFT is less resource-intensive than traditional fine-tuning, it still requires creating a training set, training the model, and evaluating its performance. This process can be time-consuming, making it slower to develop compared to simply storing embeddings in vector databases. Nonetheless, research \cite{liu2022few} indicates that combining fine-tuning with data augmentation can enhance retrieval accuracy, making it a viable option depending on the available resources, budget, and timeline.

\section{Evaluation}
    Incorporating quality assurance and testing from the early stages of product development is essential, especially when dealing with advanced technologies like LLMs and semantic retrieval, which differ significantly from traditional keyword search systems. The testing strategies, cases, and plans must be adapted to address the unique challenges posed by LLMs. Additionally, evaluating the sociotechnical aspects of the system through user surveys provides valuable insights into the effectiveness and acceptance of the technology. 

To assess the qualitative aspects of the application, we conducted pre- and post-surveys with identified pilot users. The pre-survey (Appendix \ref{sec:presurvey}), discussed earlier, was aimed at evaluating user readiness and expectations. After the pilot run, a post-survey (Appendix \ref{sec:postsurvey}) was conducted to gather feedback on the application's performance, user satisfaction, and areas for improvement.  Detailed results are tabulated in Table \ref{tab:postsurvey_analysis}

\subsection{Test Strategies}
Testing LLM-based systems necessitates a distinct approach from traditional software testing methodologies. Establishing a ground truth dataset is crucial for assessing the semantic accuracy of the generated responses. The testing process must be thorough, addressing both the generative and retrieval components of the system.

When evaluating applications developed for Retrieval-Augmented Generation (RAG) with Large Language Models (LLMs) on proprietary documents, organizations should implement a robust testing strategy to ensure the accuracy, reliability, and adherence to organizational standards. Key strategies to consider include:

\begin{itemize}
    
\item \textbf{Ground Truth Comparison}

Develop a robust ground truth dataset consisting of correct and validated responses for a set of queries relevant to the proprietary documents. This dataset serves as a benchmark to evaluate the performance of the RAG application.
Regularly compare the outputs of the RAG system against the ground truth dataset to assess the accuracy of the generated responses. Focus on key metrics such as precision, recall, and relevance.

\item \textbf{Scenario-Based Testing}

Create test scenarios that reflect real-world use cases, ensuring that the RAG application can correctly interpret and respond to complex queries within the appropriate context.

\item \textbf{Fact-Checking}

Implement automated fact-checking mechanisms to verify that the RAG system's generated content accurately reflects the source documents. One method could include embedding links to the original documents to aid in verification. 

\item \textbf{Consistency}

To assess the system's consistency in responding to similar or related queries and to prevent contradictory outputs, an external LLM—often referred to as 'LLM-as-a-judge' \cite{zheng2024judging} —can be utilized. We applied this approach by using the external LLM to evaluate the cosine similarity between responses generated for the same query to ensure consistency. Based on these evaluations, we were able to adjust parameters such as temperature, top-p, frequency penalty, and presence penalty.

\item \textbf{Performance and Scalability Testing}

\textbf{Load Testing:} Assess the application’s performance under varying loads to ensure it can handle high query volumes without degradation in response quality or speed.

\textbf{Scalability Testing:} Evaluate the system's capacity to handle increased data volume and user demand, ensuring consistent performance as the application expands. While using API-based LLM calls, we encountered rate limits during peak hours, which could degrade application performance. Frequent rate limit issues may necessitate considering alternatives such as hosted LLMs or setting token limits for API-based calls. Additionally, scalability testing will influence the choice of context summarization strategies and determine the hosting infrastructure requirements, including the need for auto-scaling.

\item \textbf{Security and Data Privacy Testing}

Evaluate the system to confirm that proprietary data is not unintentionally disclosed in generated responses. Verify that the RAG application adheres to data privacy regulations and organizational policies, especially when dealing with sensitive proprietary information. To achieve this, we involved our legal and compliance team from the onset of the project.

\item \textbf{User Experience and Feedback Integration}

\textbf{User Acceptance Testing (UAT):} Engage end-users in the testing process to collect insights on the system’s usability, relevance, and overall performance. This feedback is essential for fine-tuning the application.

\textbf{Iterative Testing with Feedback Loops:} Establish continuous feedback loops where user input drives subsequent testing and development, ensuring the application evolves to meet changing user requirements.
We shared the post-survey responses from the UAT phase and, during the pilot run, encouraged users to actively provide feedback through internal communication channels and emails. We closely tracked the key likes and dislikes about the application.

\item \textbf{Long-Term Monitoring and Evaluation}

\textbf{Post-Deployment Monitoring:} Implement ongoing monitoring of the RAG system after deployment, tracking performance metrics, user satisfaction, and compliance with governance standards.

\textbf{Periodic Audits:} Conduct regular audits of the system’s performance and compliance, revisiting and updating the ground truth dataset and testing protocols as necessary.

\item \textbf{Ethical and Explainability Testing}

\textbf{Explainability Analysis:} Assess the system’s capability to offer clear explanations for its outputs, ensuring transparency and comprehensibility for end-users. We achieved this by embedding links to the source documents.

\textbf{Ethical Considerations:} Regularly evaluate the application for adherence to ethical standards, particularly concerning the use and interpretation of proprietary data. The 'ConstitutionalChain' feature in \textsc{Langchain} \cite{Chase_LangChain_2022} can be utilized to set predefined rules that the LLM must follow.

\end{itemize}

\subsubsection{Local testing}

During this study, we explored some of the emerging solutions like \textsc{TruLens} \cite{pmlr-v176-datta22a}, \textsc{DeepEval} \cite{Ip2024confident}, \textsc{LangSmith} by \textsc{LangChain} \cite{Chase_LangChain_2022}, LlamaIndex \cite{Liu_LlamaIndex_2022} etc., which provide tools and capabilities for evaluating RAG systems. One significant challenge with LLMs is their tendency to "hallucinate" when information retrieval fails. Our study found that irrelevant LLM responses often stemmed from retrieval errors, such as when a document or its relevant chunk was not correctly retrieved in the top-n results. Recent research \cite{barnett2024seven} also identifies various points of failure in RAG systems, underscoring the importance of thorough evaluation.   

For instance, open-source frameworks like \textsc{LlamaIndex} \cite{Liu_LlamaIndex_2022} provide tools such as \textsc{Ragas} for assessing RAG performance metrics, including faithfulness, relevancy, context precision, and context recall. To use these tools, developers must create a ground truth dataset to evaluate the application's generated responses. Since such frameworks are still under active development, organizations must carefully consider their reliability before adoption. Additionally, organizations might consider employing a parallel development team to devise an automated testing strategy and methodology. Regardless of the chosen approach, it is crucial to thoroughly evaluate these metrics.

\subsubsection{Red teaming}

Red teaming, or adversarial testing, is a critical strategy for rigorously evaluating the robustness and security of a RAG system, particularly when deployed in sensitive or high-stakes environments. This approach involves deliberately challenging the system with edge cases, adversarial prompts, or scenarios designed to provoke biased, harmful, or toxic responses. The goal is to uncover potential vulnerabilities and weaknesses that might not be evident during standard testing processes.

To implement red teaming effectively, organizations can deploy a secondary LLM, known as an adversarial LLM, to generate challenging inputs. This adversarial LLM can be configured to explore the limits of the primary system's capabilities, testing its response to extreme or unexpected situations. For example, it might introduce queries that push the boundaries of the system's ethical guidelines or attempt to elicit biased or misleading information.

The insights gained from red teaming are invaluable for refining the system's safeguards, ensuring that it can handle a wide range of real-world scenarios without compromising on ethical standards or data integrity. Additionally, this process can help in identifying areas where further training or reinforcement of the LLM is necessary, contributing to the development of a more resilient and trustworthy AI system. Regular red teaming exercises, combined with iterative improvements based on the findings, are essential for maintaining the long-term security and reliability of RAG systems in any operational environment.
    
\subsubsection{User acceptance test with pilot users}
To evaluate the practical application of our system, we conducted alpha testing with pilot users, measuring both qualitative and quantitative metrics through post-survey questionnaires (Please refer Appendix \ref{sec:postsurvey}). This process helped assess the system's fitness, confidence, and projectability, providing valuable insights into its readiness for broader deployment. Please refer Table \ref{tab:postsurvey_analysis}

\begin{table}[ht]
\centering
\caption{Post-Survey Analysis Summary}
\label{tab:postsurvey_analysis}
\begin{tabular}{|p{0.3\linewidth}|p{0.3\linewidth}|p{0.3\linewidth}|}
\hline
\textbf{Question No.} & \textbf{Summary of Response} & \textbf{Percentage} \\ \hline

\textbf{1.} First impression of the
RAG application & Most users had an “Ok” first impression of the RAG application. & Majority \\ \hline

\textbf{2.} Rating of the RAG applica-
tion & Users felt the application needed improvements, with varying levels of feedback on the extent of the required improvements. & 
\begin{tabular}[c]{@{}l@{}}
- Some improvements: 52\% \\ 
- Significant improvements: 28\% \\ 
- Minor tweaks: 13\% \\ 
- Ready for use: 7\%
\end{tabular} \\ \hline

\textbf{3.} Effectiveness with context
established & The application often provided the correct answers when context was established. & Majority \\ \hline

\textbf{4.} Effectiveness without con-
text & The application rarely provided correct answers without context. & Majority \\ \hline

\textbf{5.} Helpfulness of citation
links & All users found the citation links helpful, with suggestions for improvement. & 100\% \\ \hline

\textbf{6.} Recommendation to team & Most users would recommend the application to their team, though some expressed reservations due to needed modifications. & Majority \\ \hline

\textbf{7.} Productivity improvement & Most users believed the application would moderately to significantly improve their productivity. & Majority \\ \hline

\end{tabular}
\end{table}

The post-survey results of the RAG application highlighted a mix of opinions, though feedback was generally positive compared to initial expectations. While users were initially enthusiastic about the potential of generative AI to improve efficiency, communication, and transparency, the majority rated their first impression of the RAG application as "Ok." Specifically, 52\% of users indicated that the application needed some improvements, 28\% believed significant enhancements were necessary, and only 7\% considered it ready for immediate use.

When the context was provided, users frequently received the answers they needed, but the application rarely delivered accurate results without context. Despite these challenges, all users found the citation links beneficial, with some suggesting further enhancements. Most users were inclined to recommend the application to their team, though a few noted the need for substantial modifications. Regarding productivity, most users believed the application would moderately to significantly boost their efficiency, aligning with the optimistic outlook from the pre-survey. Concerns regarding data privacy, reliability, and potential over-reliance on AI continued to resonate, similar to the pre-survey apprehensions. Overall, the application demonstrated potential, but further development is required to fully meet user expectations and realize its capabilities.

\section{AI Governance and Best Practices}
    As artificial intelligence (AI) advances, the importance of establishing robust governance frameworks and best practices becomes increasingly critical. We review previous research that highlights this necessity \cite{taeihagh2021governance} \cite{butcher2019state} \cite{dafoe2018ai}, which primarily addresses global issues concerning the development, deployment, and regulation of AI, including its intended purposes and limitations. Additionally, research has explored AI governance at the organizational level, which we use as the basis and adopt the definition stated \cite{mantymaki2022defining}:

\textit{"AI governance is a system of rules, practices, processes, and technological tools designed to ensure that an organization’s use of AI technologies aligns with its strategies, objectives, and values; complies with legal requirements; and adheres to the ethical AI principles upheld by the organization."}

Building on this definition and insights from our pilot study, we propose an AI governance model, particularly for emerging technologies such as Retrieval-Augmented Generation (RAG). RAG represents a significant advancement in AI, enabling large language models (LLMs) to produce more accurate and contextually relevant outputs by integrating external data sources that are not part of the original training data. However, the implementation of RAG in real-world scenarios presents unique challenges, particularly when transitioning from research concepts to industry applications.

In this section, we share the intersection of AI governance and the practical deployment of RAG, drawing on insights from this field study. Here, we share the importance of AI governance in ensuring that RAG implementations are not only effective but also aligned with ethical standards, regulatory requirements, and organizational goals. The study underscores the complexities involved in adopting RAG technology, including the need for stringent data management practices, transparency in AI decision-making, and the continuous monitoring of AI systems to mitigate risks such as bias, misinformation, and unintended consequences.

From what we have learned, we provide a foundation for developing best practices that organizations can adopt to navigate the challenges of implementing RAG. These best practices address key areas such as data governance, system design, user engagement, and ongoing evaluation, all of which are essential for maximizing the benefits of RAG while minimizing potential pitfalls. By sharing these insights, we aim to contribute to the broader discourse on AI governance and offer actionable guidance for organizations looking to harness the power of RAG in a responsible and sustainable manner.

\subsection{AI Governance}
\begin{figure}
\includegraphics[width=\textwidth]{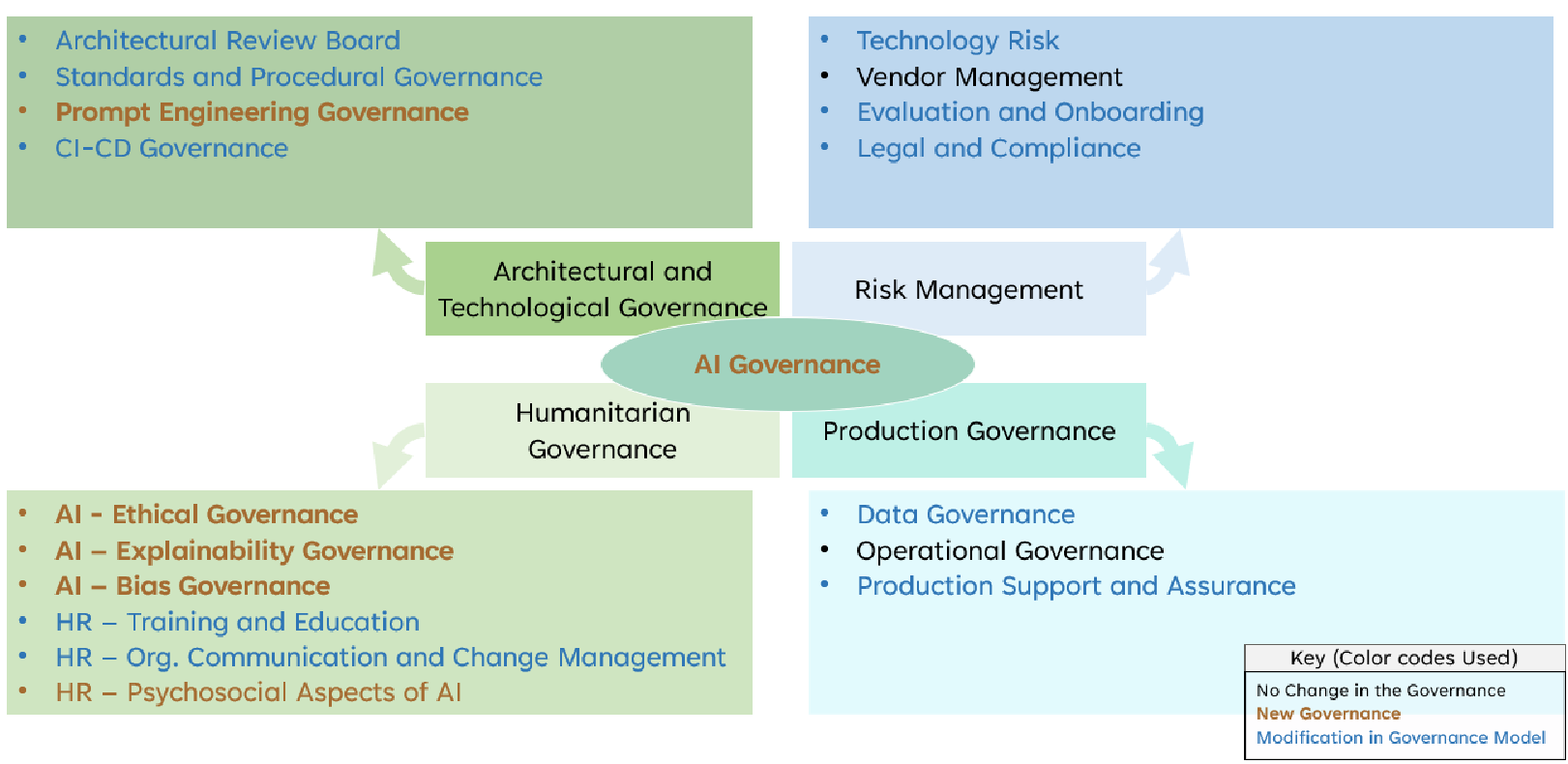}
\caption{AI Governance Model} \label{fig3}
\end{figure}

In our field study, we developed an AI governance model designed to address the unique challenges posed by AI systems, particularly those involving Retrieval-Augmented Generation (RAG). Our proposed governance model includes both enhancements to existing governance frameworks and the introduction of novel governance practices specifically tailored to AI technologies. The model is divided into four broad categories: Architectural and Technological Governance, Risk Governance, Humanitarian Governance, and Production Governance. (Please refer Figure \ref{fig3}) Each category addresses critical aspects of AI system implementation, ensuring that organizations can effectively manage the complexities of AI deployment while adhering to ethical, legal, and operational standards.

\subsubsection{Architectural and Technological Governance} 
Given the distinct differences between AI systems and traditional information systems, significant updates to architectural and technological governance are required: 

\textbf{Architectural Review Board} For AI implementations, the Architectural Review Board must establish processes to differentiate between cloud-based and on-premise infrastructure. It is essential to make informed decisions regarding the use of on-premise AI models versus API-based AI models, considering factors such as data security, performance, and scalability.

\textbf{Standard and Procedural Governance} This aspect covers the development and enforcement of standards, guidelines, and best practices for integrating AI capabilities within both existing and new applications. 

\textbf{Prompt Engineering Governance} The effectiveness of an AI system, particularly those leveraging large language models (LLMs), heavily depends on the clarity, structure, and flexibility of the prompts used to guide the model's responses. With LLMs, prompt engineering lies at the heart of the system. Therefore, it is important to discuss this new governance for prompt engineering in more detail to ensure consistency, reusability, and maintainability of these prompts.

Structuring the prompt templates through standardization is crucial for maintaining consistency across different applications and teams. These templates should be designed to be adaptable, allowing for easy customization while maintaining a core structure that ensures reliability. In addition, using a modular design, such as breaking the prompt into sections for prefix, context, instructions, and output format, allows teams to reuse and repurpose different sections of the prompt across various scenarios, enhancing efficiency and reducing the need for constant re-engineering.

To enable reusability and flexibility, we propose establishing a library of prompt templates, thus enablimg teams to access a repository of well-tested prompts that can be adapted for various use cases. This reduces development time and ensures that best practices are consistently applied across the organization. To manage the evolution of prompt templates, version control should be implemented. This allows teams to track changes, revert to previous versions if necessary, and ensure that updates are systematically applied across all relevant systems. Version control also facilitates collaboration, enabling multiple teams to contribute to and refine prompt templates over time.
While standardization is important, prompts must also be flexible enough to adapt to specific contexts or scenarios. Governance should include guidelines on how to modify prompts for different applications while maintaining the core principles of clarity, accuracy, and efficiency.

For maintenance and continuous improvement, prompt templates should undergo regular audits to ensure they remain effective and aligned with the latest developments in AI technology and organizational needs. These audits should assess the performance of prompts, identifying areas for improvement and updating templates accordingly. Establishing feedback mechanisms is crucial for the continuous improvement of prompt engineering. Teams should collect and analyze feedback from end-users and developers to refine prompts and address any issues that arise during deployment. This iterative process helps to keep prompts relevant and effective in changing environments.
In addition, comprehensive training and documentation are essential for ensuring that all team members understand how to create, use, and maintain prompt templates. Documentation should cover the principles of prompt engineering, provide examples of best practices, and offer guidance on troubleshooting common issues.

Finally, effective prompt engineering governance requires input from various disciplines, including AI specialists, domain experts, and end-users. This cross-functional collaboration approach ensures that prompts are not only technically sound but also aligned with the specific needs of the application and the users it serves.
Prompt engineering governance should not be the sole responsibility of a single team. Instead, it should be a shared responsibility across the organization, with clear roles and responsibilities assigned to different teams. This approach ensures that prompt engineering remains a priority and that best practices are consistently applied across all projects.

\textbf{CI-CD (Continuous Improvement / Continuous Development) Governance} The rapidly changing landscape of AI technology necessitates a reevaluation of existing CI-CD practices. Organizations must adjust their frequency of data and code updates to stay aligned with the latest advancements in AI vendor technologies. Additionally, changes in laws and regulations affecting AI implementations must be incorporated into these processes. 

\subsubsection{Risk Governance}
Risk governance provides oversight to assess and mitigate the potential risks associated with AI implementations, ensuring that they do not adversely impact the organization’s reputation or operational integrity: 

\textbf{Technology Risk} Due to the emergent nature of AI, coupled with its rapid evolution, AI-based applications face heightened risks. These risks must be regularly evaluated and addressed through comprehensive risk management strategies, ensuring that the AI systems remain robust and reliable over time. 

\textbf{Vendor Management} While the fundamentals of vendor management remain largely unchanged, the evaluation and onboarding of AI vendors require careful consideration. Organizations must assess whether new AI models utilize data responsibly, comply with enterprise contracts, and meet security standards before integration. 

\textbf{Evaluation and On-boarding} The evaluation of newer models and whether those utilize the data sent for training has to be considered before on-boarding the new AI models and technologies. Evaluating enterprise contracts and security compliance is also part of the evaluation and onboarding governance. 

\textbf{Legal and Compliance} Keeping abreast of the dynamic regulatory environment surrounding AI is crucial. Organizations must establish governance frameworks that ensure compliance with evolving legal requirements and guide the implementation of AI systems in line with these standards.

\subsubsection{Humanitarian Governance}
Humanitarian governance focuses on the socio-technological impacts of AI systems, particularly on human-AI interactions, ethics, and organizational culture:

\textbf{Ethical Governance} AI implementations must adhere to established ethical norms. This governance ensures that AI systems are developed and deployed in a manner that respects human rights, fairness, and transparency.

\textbf{Explainability Governance} To foster trust and accountability, it is imperative that AI systems are explainable. This governance sets the required level of explainability for AI systems within the organization and mandates adherence to these standards across all AI-based solutions. 

\textbf{Bias Governance} Addressing biases in AI systems is critical. This governance requires the analysis of potential biases in AI implementations and the development of mitigation strategies to ensure fairness and equity in AI-driven decisions.

\textbf{Training and Education} While training is an existing governance model, AI systems demand specialized training for teams involved in their development and use. This includes educating allied teams and customers on how to effectively interact with AI systems.

\textbf{Organizational Communication and Change Management} Existing governance models for communication and change management must be adapted to meet the specific requirements of AI systems. 

\textbf{Psycho-social Aspects of AI} This includes addressing the psycho-social aspects of AI, such as managing staff and user anxieties through targeted training and support programs, thereby enhancing the adoption of AI technologies. 

\subsubsection{Production Governance}
Production governance ensures that AI systems are managed effectively once deployed, encompassing monitoring, disaster recovery, business continuity, incident management production support, etc. 

\textbf{Data Governance} Data governance for AI systems must be restructured to account for the dynamic nature of data that can modify AI models and outputs. This involves implementing stringent data quality controls and ensuring that the data feeding into AI models is reliable and accurate. 

\textbf{Operational Governance} While operational governance practices such as staff backups, rotation schedules, and incident management remain largely applicable to AI systems, they must be fine-tuned to accommodate the specific operational needs of AI technologies. 

\textbf{Production Support and Assurance} AI-based systems require specialized production support. Organizations must equip their production support teams with the necessary training and resources to handle the unique challenges posed by AI systems, ensuring that they can provide effective and timely support.

This AI governance model is based on our learnings from this field study and is designed to provide a comprehensive framework that organizations can adopt to manage the complexities of AI deployment. By integrating these governance practices, organizations can ensure that their AI systems are not only effective but also ethical, compliant, and resilient in the face of evolving technological and regulatory landscapes.

\subsection{Best Practices}
In our field study, we experimented with various prompt patterns \cite{white2023prompt} and explored different aspects of Retrieval-Augmented Generation (RAG) to develop best practices for building LLM-based systems. The following recommendations are grounded in lessons learned from our build-test cycles and are designed to address key areas such as data governance, system design, user engagement, and ongoing evaluation.

\subsubsection{Prompt Engineering}

\textbf{Prompt Structure:} RAG allows LLMs to stay focused and minimize hallucinations by providing clear context through the prompt. We developed structured templates that enhance readability and flexibility during development. These templates also segment instructions into distinct sections, which is crucial for maintaining clarity. Incorporating a prefix section in the prompt helps define the LLM's persona and provides structural information, ensuring the model understands the organization of the prompt. Custom begin and end tags (e.g., <|im-start|>, <|im-end|>) facilitate easy parsing of the prompt's sections. 

\textbf{Prompt Writing:} Clarity and specificity in prompt instructions are vital for guiding the LLM effectively. For instance, instructions like "If there is not enough information, say 'I don't know'" or "Be brief in your answers" help the model deliver more accurate responses. Capitalizing key directives (e.g., "DO NOT," "ONLY") and numbering instructions can further emphasize critical points. When expecting output in a specific format (e.g., JSON), providing an example is beneficial. Additionally, using the function-calling feature of LLMs, such as in ChatGPT, can help ensure the output follows a particular pattern.

\subsubsection{Data Governance}

\textbf{Vector Databases:} Experimenting with various vector databases and search mechanisms is crucial for optimizing RAG systems. We tested databases like Chroma, Pinecone, and Elastic Relevance Search Engine and compared different search strategies, including exact-match and semantic search, or a combination of both. Proper data governance involves selecting and managing these databases to ensure data accuracy, consistency, and privacy throughout the retrieval process. Establishing clear guidelines for data handling, storage, and retrieval is essential for maintaining the integrity of the information used by RAG systems.

\subsubsection{System Design}

\textbf{Modular Architecture:} Given the rapidly evolving nature of AI technologies, designing the application in modular, decoupled components is advisable. This approach allows for easier updates and integration of new features without disrupting the entire system. Thorough planning of the architecture before building the prototype is essential. Extensive logging should be implemented to track which documents were retrieved for specific user queries and what responses were generated. This not only aids in debugging but also supports ongoing evaluation and improvement of the system.

\textbf{LLM Considerations:} Selecting the right AI provider is critical. Enterprise accounts offer more reliable API access and support, which is particularly important for large-scale applications. It's beneficial to compare the capabilities of cloud-based LLMs (e.g., those offered by Google Vertex AI or Microsoft Azure OpenAI) with locally hosted, fine-tuned models on proprietary data. Each option has distinct advantages, depending on the business case, and should be evaluated in terms of scalability, cost, and data security.

\subsubsection{User Engagement}

\textbf{Human Factors:} Engaging end users early in the development process is key to creating a system that meets their needs. Request users to provide a set of frequently used questions for quality testing and unit testing. This approach helps mitigate developers' bias and avoids tunnel vision, ensuring the system is tested against real-world scenarios that users will encounter. Continuous feedback from users should be integrated into the development cycle to refine the system's performance.

\subsubsection{Ongoing Evaluation}

\textbf{Continuous Monitoring and Feedback:} Implementing a robust system for ongoing evaluation is crucial for maintaining the effectiveness and relevance of the RAG system. This includes regular monitoring of the system's performance, user satisfaction, and compliance with data governance policies. Feedback loops should be established to capture insights from users and operational data, allowing for iterative improvements. Regular updates to the system based on these evaluations will help ensure that it continues to meet evolving user needs and technological advancements.

By following these best practices that we contribute based on the insights from our field study, organizations can successfully implement RAG-based solutions using LLMs, ensuring that the technology is applied in a way that is effective, responsible, and aligned with industry standards and user expectations.

\section{Discussion and Implications}
    This research and field study uncovered several technological, enterprise, and human factors that must be addressed as Retrieval-Augmented Generation (RAG) systems continue to evolve. From a technological perspective, one major challenge encountered was the limitations associated with using external AI models like \textsc{ChatGPT} via APIs, such as rate limits, delayed responses, and token constraints. These limitations directly impact the performance and scalability of RAG applications, requiring careful consideration of model selection, token management, and response optimization.

On the enterprise front, the use of external APIs for large language models (LLMs) necessitates a rigorous assessment of the proprietary information being exposed to these models, particularly in sensitive contexts where data security is paramount. Gaining stakeholder buy-in and securing budget allocations are critical steps in developing systems based on emerging technologies. This highlights the importance of robust AI governance frameworks, which we addressed in our study by proposing governance models that account for the unique requirements of AI systems, including prompt engineering, risk management, and legal compliance.

Our pilot phase underscored the importance of managing user expectations for the successful deployment of AI-based applications. We found that user training is critical, particularly in question-and-answer interactions, where poorly framed questions can result in inaccurate or misleading responses. By educating users on how to interact effectively with AI systems, many of these issues can be mitigated, leading to improved overall system performance and user satisfaction.

In addressing the challenges we faced, we encountered issues similar to those documented by Hevner et al. \cite{hevner2023research} in their research on the design of human-artificial intelligence systems (HAIS). These challenges included selecting appropriate technologies amid legal and compliance constraints, managing the complexity of integrating diverse data sources, and evaluating the effectiveness of LLM-based applications. Each of these issues required careful consideration and iterative problem-solving, such as modular design, user feedback through pre-surveys, and the strategic use of tools like \textsc{TruLens} \cite{pmlr-v176-datta22a}, \textsc{DeepEval} \cite{Ip2024confident}, \textsc{LangSmith} by \textsc{LangChain} \cite{Chase_LangChain_2022}, etc., for evaluation.

In addition, the implementation of Retrieval-Augmented Generation (RAG) systems in real-world scenarios necessitates a robust AI governance framework to ensure ethical, legal, and operational integrity. As AI technologies evolve, especially with advancements like RAG, governance models must adapt to address new challenges, including architectural design, risk management, and the socio-technical impacts on human-AI interactions.

Our field study insights laid the groundwork for developing the proposed AI governance model tailored specifically to RAG systems, with a focus on key areas such as prompt engineering, data governance, and continuous evaluation. This governance model emphasizes the importance of maintaining transparency, ensuring ethical AI deployment, and managing risks associated with the rapid evolution of AI technologies. By adopting these governance practices, organizations can deploy RAG systems effectively while adhering to compliance standards and mitigating potential risks such as bias or misinformation. Integrating these insights into AI governance contributes to the broader discourse on responsible AI deployment and offers a practical framework for organizations aiming to harness the potential of RAG systems. 

Given the complex challenges revealed by our field study, it is essential for industry and academia to collaborate more closely to advance AI technologies in a manner that is both innovative and responsible. By sharing the real-world difficulties encountered during implementations, research in the field of LLM and RAG systems can be significantly accelerated. This collaboration will not only refine best practices and governance models but also foster the development of robust, reliable, and compliant AI solutions that can be widely adopted across various industries.

This study enhances the growing body of knowledge on RAG systems by providing a comprehensive analysis of the technical, organizational, and human challenges involved in their implementation. Our findings highlight the importance of a holistic approach that integrates AI governance, user engagement, and continuous evaluation, offering a roadmap for organizations looking to adopt RAG-based solutions in a dynamic technological landscape.

\section{Conclusion}
    This study has highlighted the key challenges and opportunities associated with implementing Retrieval-Augmented Generation (RAG) systems using large language models (LLMs). While RAG is emerging as a significant application of LLMs, our findings indicate that it remains in its early stages, with many organizations still navigating the complexities of its deployment. The rapid evolution of underlying technologies adds to the challenge, making it a continuously moving target.

In addition to addressing technical challenges, our research has also contributed to AI governance frameworks and best practices essential for successfully integrating RAG systems. We proposed governance models that account for the unique demands of AI systems, emphasizing the need for structured prompt engineering, comprehensive risk management, ethical considerations, and ongoing system evaluation. These governance recommendations provide organizations with a robust foundation for ensuring that their AI implementations are effective and compliant with emerging standards and regulations.

By sharing the lessons learned and best practices from our field study, we aim to offer a practical framework that organizations can use to overcome these obstacles. The insights gained from our research offer valuable guidance for both technical and governance aspects for those embarking on similar ventures, helping to ensure that the challenges of integrating RAG into production environments can be more effectively addressed. We hope that these contributions will support the broader adoption and refinement of RAG systems, advancing both industry practice and academic research in this area.

\bibliographystyle{unsrt}  
\bibliography{references}  

\appendix
 \section{Appendix}
\label{sec:funcnonfunc}
\subsection{Functional Requirements}
\begin{enumerate}
    \item Natural Language Understanding (NLU): The application should be able to understand and interpret user inputs accurately, even when the language used is informal or contains typos.
    \item Contextual Understanding: The application should maintain the context of a conversation, allowing it to provide relevant responses and follow up on previous interactions.
    \item Multi-Platform Compatibility: The application architecture should be compatible with various platforms, including websites, mobile apps, and social media platforms.
    \item Integration with Existing Systems: The application should be able to integrate with existing enterprise systems to access and provide relevant information.
    \item It should support various interfaces to various document formats, such as
    \begin{enumerate}
        \item Written documents (PDFs, Word, Excel, etc.)
        \item Organization’s internal communicator data.
        \item Integrate various existing databases.
    \end{enumerate}
    \item Self-Learning: The application should consume individual documents or documents placed in a specific directory.
    \item When new documents are placed into a specified directory, the application shall check on an interval time for new documents to ingest and add to the indexes.
    \item User Authentication: The application should be able to authenticate users to provide personalized services and handle sensitive information securely.
    \item User sessions should be separated so that one user’s questions and answers are not mixed between conversations.
    \item Authentication should be done as a Single Sign On (SSO) with the organization’s Active Directory
    \item The application should support various authorization levels at individual and group levels.
    \item Response Time: The application should respond to user inputs within a reasonable time frame, ensuring a smooth and efficient user experience.
\end{enumerate}

\subsection{Non-Functional Requirements}
\begin{enumerate}
    \item Scalability: The application should be able to handle many simultaneous conversations without any significant degradation in performance.
    \item Security: The application should comply with all relevant data protection and privacy regulations. User data should be encrypted and stored securely.
    \item Reliability: The application should be reliable, with a high uptime and minimal errors or failures.
    \item Usability: The application should be easy to use, with a user-friendly interface and clear, understandable responses.
    \item Maintainability: The application should be easy to update and maintain, allowing for continuous improvement and the addition of new features.
\end{enumerate}

\subsection{Technical Requirements}
\begin{enumerate}
    \item Vector Database – Chroma DB
    \item UI - Self-hosted React.js Frontend
    \item API – Flask API
    \item Software language and toolkits – Python, Langchain
\end{enumerate}

\subsection{Stakeholder Requirements}
\begin{enumerate}
    \item As a user, I want the application to understand my queries, so that I can get accurate and relevant responses.
    \item As a user, I want the application to remember previous interactions so I don't have to repeat information.
    \item As a user, I want the application to provide citations to document sources so I can confirm and expand upon the generated information.
    \item As a business owner, I want the application to integrate with my existing systems to access and provide up-to-date information.
\end{enumerate}

\subsection{Acceptance Criteria}
The AI application will be considered complete when it meets all the functional and non-functional requirements outlined and successfully fulfills the stakeholder requirements.

\subsection{Stakeholders}
\begin{enumerate}
    \item Solution Architect team
    \item Development team
    \item Quality Assurance team
    \item Performance testing team
    \item Technology risk management team
    \item Enterprise production support teams
    \item Network and infrastructure teams
    \item Shared services teams
    \item Customer support end users
    \item Legal, compliance, and regulatory teams
    \item Executive leadership team, including heads of engineering, strategy, marketing, training, and documentation organizations within the company.
\end{enumerate}

\subsection{Measures of effectiveness}
\subsubsection{Technical}
\begin{itemize}
\item Response should be relevant to the question asked.
\item Response should be accurate, factual, and generated based on the existing documentation.
\item Each response should provide citation links to the documentation from which the response was generated.
\item Responses should be consistent every time the same question is asked. The response can be aesthetically different (worded differently), but the content should be the same.
\end{itemize}
\subsubsection{Human}
\begin{itemize}
\item The system should be easy to use, user-friendly, and intuitive.
\item The system should be reliable and robust so users can use it without double-checking the cited documents.
\item The users should be provided with a consistent interface such that any changes to the back-end technology should be agnostic to the user.
\end{itemize}

\section{Pre-survey Questionnaire}
\label{sec:presurvey}
\begin{enumerate}
    \item The first few questions are related to personal information such as name, team, age, department, years of service in the organization, designation, etc.
    \item Have you used a generative AI application?
    \item How do you think using generative AI would help you and your team to be more efficient?
    \item How do you think generative AI can help with cross-functional teams?
    \item What are your expectations from the Information Retrieval System using Generative AI?
    \item Do you have any concerns with using generative AI applications?
    \item What data or information do you expect the generative AI application to give?
    \item Any additional inputs, comments, or notes.
\end{enumerate}

\subsection{Analysis from the pre-survey}
\begin{enumerate}
    \item Question 1: This data is redacted for privacy reasons. The team of pilot users was across various departments in the organization.
    \item Question 2: 54\% of users have some prior experience using generative AI at various levels of expertise.
    \item Question 3: Users are looking for overall accuracy and relevance of retrieved information, help with troubleshooting, fast access to information, specifics about products and releases, etc. About 8\% of the users were not sure if AI can make them or their teams more efficient.
    \item Question 4: Users are looking to reduce cross-functional team communications and meetings, as with the AI application, everyone would have quick access to everything. The users are expecting a significant increase in efficiency. About 8\% of users are “unsure but excited to find out.”
    \item Question 5: Most users were excited about the new application. Some even believed in endless possibilities, and some were excited about how it would bring the entire organization together and “all-inclusive” and increase transparency. About 20\% of the users did not know what to expect from such an application.
    \item Question 6: Concerns were about phrasing questions to the AI and some known issues with the current AI models available in the market, such as taking questions out of context, accuracy, and reliability of responses, data privacy and security, too much dependency on the AI application, and even such a tool replacing human jobs. About 41\% of the users did not have any concerns.
    \item Question 7: Most users wanted data to be retrieved across various enterprise systems in addition to documents in different formats. About 8\% of users were not sure.
    \item Question 8: Few comments in notes. About 25\% of users were excited and grateful to be included in the pilot-run program.
\end{enumerate}

\section{Post Survey Questionnaire}
\label{sec:postsurvey}
\begin{enumerate}
    \item What is your first impression of the RAG application in use?
    \begin{enumerate}
        \item Bad
        \item Ok
        \item Good
    \end{enumerate}
    \item How would you rate the RAG application?
    \begin{enumerate}
        \item Bad – Discontinue further development efforts
        \item Not Bad – Needs a lot of improvements
        \item OK – Needs some improvements
        \item Good – Needs some tweaks, but overall good
        \item Excellent – It is remarkable, can use it right now
    \end{enumerate}
    \item How often did the RAG application provide the answers you needed when the context was established?
    \begin{enumerate}
        \item Never
        \item Rarely
        \item Sometimes
        \item Often
        \item Every time
    \end{enumerate}
    \item How often did the RAG application provide the answers you needed when the context was NOT established?
    \begin{enumerate}
        \item Never
        \item Rarely
        \item Sometimes
        \item Often
        \item Every time
    \end{enumerate}
    \item Are the citation links helpful? Yes/No/Maybe
    \item Would you recommend the RAG application to your team? Yes/No/Maybe
    \item Do you think the RAG application would improve your productivity?
    \begin{enumerate}
        \item Not at all
        \item Less than 30\% - Slightly
        \item 30 - 60\% – Moderately
        \item 60 - 100\% – Significantly
    \end{enumerate}
    \item What are the top 3 items you want improved?
    \item What feature/s would you like to see added to generative AI application 2.0?
    \item Do you have any concerns about using the RAG application? If so, what?
\end{enumerate}

\subsection{Analysis from the post-survey}
\begin{enumerate}
    \item Question 1: Most users chose the option B for ok.
    \item Question 2: 52\% of users thought the application needed some improvements. 28\% of users selected the option that the application needed a lot of improvements. 13\% of users selected the option that the application needed some minor tweaks. 7\% of users selected the option that the application was ready for use.
    \item Question 3: Most users selected option C or D, which meant that when the context was provided, the application gave results in some or most cases.
    \item Question 4: Most users selected option B, which meant that with less or no context provided, the application rarely gave the answers correctly.
    \item Question 5: All users thought the citations were helpful. We also received some good feedback on improving the citations by highlighting the text or showing the user the exact page or paragraph from the document.
    \item Question 6: Most users selected Yes, and some selected Maybe for recommending the application to their team, but a few did select No for this answer. The reason for selecting No was that they thought the application needed a lot of modifications before it was ready for use.
    \item Question 7: Most users selected options C and D, and some selected B for recommending the application to their team, but a few did select No for this answer. The reason for this, in most users' opinion, was that the application needed a lot of modifications before it was ready for use.
    \item Question 8, 9, 10: The answers for these three questions are associated closely with the proprietary application, its functioning, and the context of this particular field study, and hence, this data has been redacted from the survey analysis for privacy reasons.
\end{enumerate}

\end{document}